\documentclass[journal,twoside]{IEEEtran}
\usepackage{cite}
\usepackage{amsmath,amssymb,amsfonts}
\usepackage{algorithmic}
\usepackage{graphicx}
\usepackage{textcomp}
\usepackage{booktabs}
\usepackage{url}

\usepackage{wrapfig,colortbl}
\usepackage{pdfpages}
\usepackage{soul}

\begin{document}
\title{Waveform-Specific Performance of Deep Learning-Based Super-Resolution for Ultrasound Contrast Imaging}
\author{
    Rienk Zorgdrager,
    Nathan Blanken,
    Jelmer M. Wolterink, \IEEEmembership{Member, IEEE},
    Michel Versluis, \IEEEmembership{Member, IEEE}, and
    Guillaume Lajoinie, \IEEEmembership{Member, IEEE}
    \thanks{\copyright 2022 IEEE.  Personal use of this material is permitted.  Permission from IEEE must be obtained for all other uses, in any current or future media, including reprinting/republishing this material for advertising or promotional purposes, creating new collective works, for resale or redistribution to servers or lists, or reuse of any copyrighted component of this work in other works.}
    \thanks{Rienk Zorgdrager and Guillaume Lajoinie acknowledge funding from the HORIZON.1.1 Programme of the European Research Council for the Super-FALCON project, Grant agreement ID: 101076844 (DOI: 10.3030/101076844).
    Nathan Blanken, Jelmer M. Wolterink, Michel Versluis, and Guillaume Lajoinie acknowledge funding from the 4TU Precision Medicine program supported by High Tech for a Sustainable Future, a framework commissioned by the four Universities of Technology of the Netherlands. (Corresponding author: Guillaume Lajoinie.)}
    \thanks{Rienk Zorgdrager, Nathan Blanken, Michel Versluis, and Guillaume Lajoinie are with the Physics of Fluids Group, Technical Medical (TechMed) Centre, University of Twente, Enschede, The Netherlands (e-mail: r.c.zorgdrager@utwente.nl; n.blanken@utwente.nl; m.versluis@utwente.nl; g.p.r.lajoinie@utwente.nl).}
    \thanks{Jelmer M. Wolterink is with the Department of Applied Mathematics, Technical Medical (TechMed) Centre, University of Twente, Enschede, The Netherlands (e-mail: j.m.wolterink@utwente.nl).}
    \thanks{The code and dataset used in this work are available at respectively \protect\url{https://github.com/MIAGroupUT/super-resolution-waveforms} and \protect\url{https://doi.org/10.4121/cc1c073d-23bf-4a1e-b9f4-9f878c95722d}.}
}

\maketitle

\begin{abstract}
Resolving arterial flows is essential for understanding cardiovascular pathologies, improving diagnosis, and monitoring patient condition. Ultrasound contrast imaging uses microbubbles to enhance the scattering of the blood pool, allowing for real-time visualization of blood flow. Recent developments in vector flow imaging further expand the imaging capabilities of ultrasound by temporally resolving fast arterial flow. The next obstacle to overcome is the lack of spatial resolution. 
Super-resolved ultrasound images can be obtained by deconvolving radiofrequency (RF) signals before beamforming, breaking the link between resolution and pulse duration. Convolutional neural networks (CNNs) can be trained to locally estimate the deconvolution kernel and consequently super-localize the microbubbles directly within the RF signal. However, microbubble contrast is highly nonlinear, and the potential of CNNs in microbubble localization has not yet been fully exploited. Assessing deep learning-based deconvolution performance for non-trivial imaging pulses is therefore essential for successful translation to a practical setting, where the signal-to-noise ratio is limited, and transmission schemes should comply with safety guidelines. 
In this study, we train CNNs to deconvolve RF signals and localize the microbubbles driven by harmonic pulses, chirps, or delay-encoded pulse trains. Furthermore, we discuss potential hurdles for in-vitro and in-vivo super-resolution by presenting preliminary experimental results. We find that, whereas the CNNs can accurately localize microbubbles for all pulses, a short imaging pulse offers the best performance in noise-free conditions. However, chirps offer a comparable performance without noise, but are more robust to noise and outperform all other pulses in low-signal-to-noise ratio conditions.
\end{abstract}


\section{Introduction}

\IEEEPARstart{B}{lood} flow quantification is essential for diagnosing and monitoring cardiovascular diseases because of the intrinsic link between blood flow anomalies and cardiovascular pathology.
In the current clinical workflows, ultrasound Doppler has become the standard for blood flow measurement~\cite{Christopher1996} because of its relatively low cost and ability to image flow in real-time.
Ultrasound Doppler, however, has shortcomings in both resolution and accuracy. Its spatial resolution is limited by the wavelength of the ultrasound waves. In addition, the temporal resolution of conventional Doppler is relatively low, since its processing typically requires a series of acquisitions. Furthermore, it can only measure velocities in the direction of the beam, limiting its accuracy for vessels running parallel to the transducer surface and making velocity estimations prone to errors~\cite{Hansen2017,Steinman2001}.
In the past decades, preclinical research efforts have tried to overcome these issues and aimed to produce blood flow images using speckle tracking rather than Doppler shifts. More specifically, these vector flow imaging techniques use local image cross-correlation to compute velocity fields without the angle limitations of Doppler \cite{Hansen2017, Engelhard2021, Zhou2019, Engelhard2022, Bakker2024}.

However, blood is a poor ultrasound scatterer and provides a low signal-to-noise ratio (SNR), thereby limiting the accuracy of these algorithms for blood flow imaging.
The low SNR in ultrasound vascular imaging has led to the introduction of now widespread ultrasound contrast agents. They consist of a suspension of microbubbles stabilized by a viscoelastic shell~\cite{Stride2020}. 
Once injected into the bloodstream, these bubbles enhance the scattering from the blood pool~\cite{Quaia2005}. Contrast-enhanced ultrasound imaging allows for dynamical assessment of organ microvascular perfusion~\cite{Quaia2010} and has been proposed as a technique for imaging angiogenesis \cite{Mischi2012} and evaluating malignancy grades in cancer~\cite{Masumoto2016}.
Vector flow imaging techniques in particular benefit from this increase in speckle visibility, as it increases the reliability of local cross-correlation that is needed for accurate flow velocity estimates~\cite{Engelhard2021,Zhou2019,Engelhard2022}.

While the combination of flow vector imaging techniques and microbubble contrast enhancement improves clinical flow estimates, the conventional B-mode images that are used to track speckle are diffraction-limited and as a result the resolution, and ultimately the performance of vector flow imaging algorithms, is limited by the duration of the transmit pulses. Short imaging pulses thus provide optimal B-mode resolution but this comes at the cost of a low SNR. 
This resolution limitation further hinders the adoption of these novel techniques through clinical workflows as it prevents conventional ultrasound techniques from resolving small-scale fluid mechanics. For example, wall shear stress, stagnation points, and vortical structures are known to play a key role in cardiovascular lesion development~\cite{Warboys2011, Chiu2011}, but cannot be reliably resolved.
Diagnosing and monitoring cardiovascular diseases in which such phenomena are suspected to play a key role thus requires a major improvement in the resolution of ultrasound flow imaging.

Recent developments in ultrasound localization microscopy (ULM) can overcome the diffraction limit and provide high-resolution images of capillary flows.
ULM is based on isolating non-overlapping point spread functions in B-mode images of sparse microbubble suspensions~\cite{Couture2018}.
Tracking the centroids of these point spread functions then provides sub-pulse length resolution. However, due to the sparsity requirement, ULM requires long acquisition times, and while generating large datasets also substantial processing times.
More recent investigations have been able to leverage the capabilities of deep learning to track microbubbles with lower computational cost, increased speed, and higher localization precision in microvasculature~\cite{Luan2023}. 
However, the sparsity requirement in ULM remains, and with it long acquisition times~\cite{Hingot2019}, which hinders the real-time imaging of flows. Furthermore, the long acquisition times prevent insight into time-dependent phenomena, which are particularly present in arterial pulsatile hemodynamics.

To enable sub-pulse length resolution with the advantages of real-time imaging at higher bubble densities, Blanken~\textit{et al.} have previously proposed a strategy to localize microbubbles using a convolutional neural network (CNN) to directly deconvolve the RF signals before beamforming. This approach does not rely on separating isolated point spread functions and is free from artifacts introduced during beamforming~\cite{Blanken2022}. Furthermore, working directly on the RF data allows for exploiting the full information content of the raw data, such as nonlinearities, which are largely lost during image reconstruction.

The CNN trained by Blanken~\textit{et al.}~\cite{Blanken2022} learns to estimate microbubble location by deconvolving a synthetic RF signal affected by a set of nonlinearities and spatial variations, including the driven, nonlinear microbubble response, nonlinear propagation within the transmit field, and a relevant range of ultrasound pressures.

A related problem in conventional ultrasound imaging is the recovery of image resolution upon using long but carefully designed pulses: the so-called coded excitation pulses. These pulses are constructed using waveforms that have a narrow autocorrelation, such as chirps~\cite{Sun2007}, second-order ultrasound field (SURF) pulses~\cite{Hansen2009}, and cascaded waves~\cite{Bakker2024}. Coded excitation pulses are typically longer than the usual imaging pulses, which would a priori lead to a lower imaging resolution. However, when using an appropriate decoding strategy on the RF signals prior to beamforming, these pulses can be used to increase the SNR without sacrificing resolution~\cite{Borsboom2005,Bottenus2023,ODonnell1992}.

Coded excitation pulses, and even more so, CNN-based deconvolution, break with the conventional paradigm that links pulse length and imaging resolution. To bridge the large gap between idealized synthetic data and clinical implementation of deep learning-based deconvolution, it is essential to evaluate how the design of the ultrasound pulse can be used to maximize performance and potential of CNNs for super-resolution. 
In this study, we thus depart from this resolution-pulse length paradigm and investigate whether the performance of CNN-based deconvolution can be improved using imaging pulses that bear a fundamental interest in ultrasound imaging, beyond the traditional, short, imaging pulses. 
We investigate the 1D microbubble localization performance of networks trained on simulations that emulate harmonic ultrasound pulses of different durations and center frequencies, as well as coded excitation pulses with and without decoding. The former is motivated by the known dependence of bubble nonlinearity on the driving pulse length, and the latter exploit the concept of coded excitation in ultrasound imaging. Since SNR is a critical quantity in ultrasound (contrast) imaging, we further evaluate the potential of our deep learning-based deconvolution technique for increasing levels of random noise in the RF signals. Furthermore, we perform a preliminary evaluation of the technique on experimental data.

\section{Methods}
\label{sec: Methods}

\begin{figure*}
    \centering
    \includegraphics{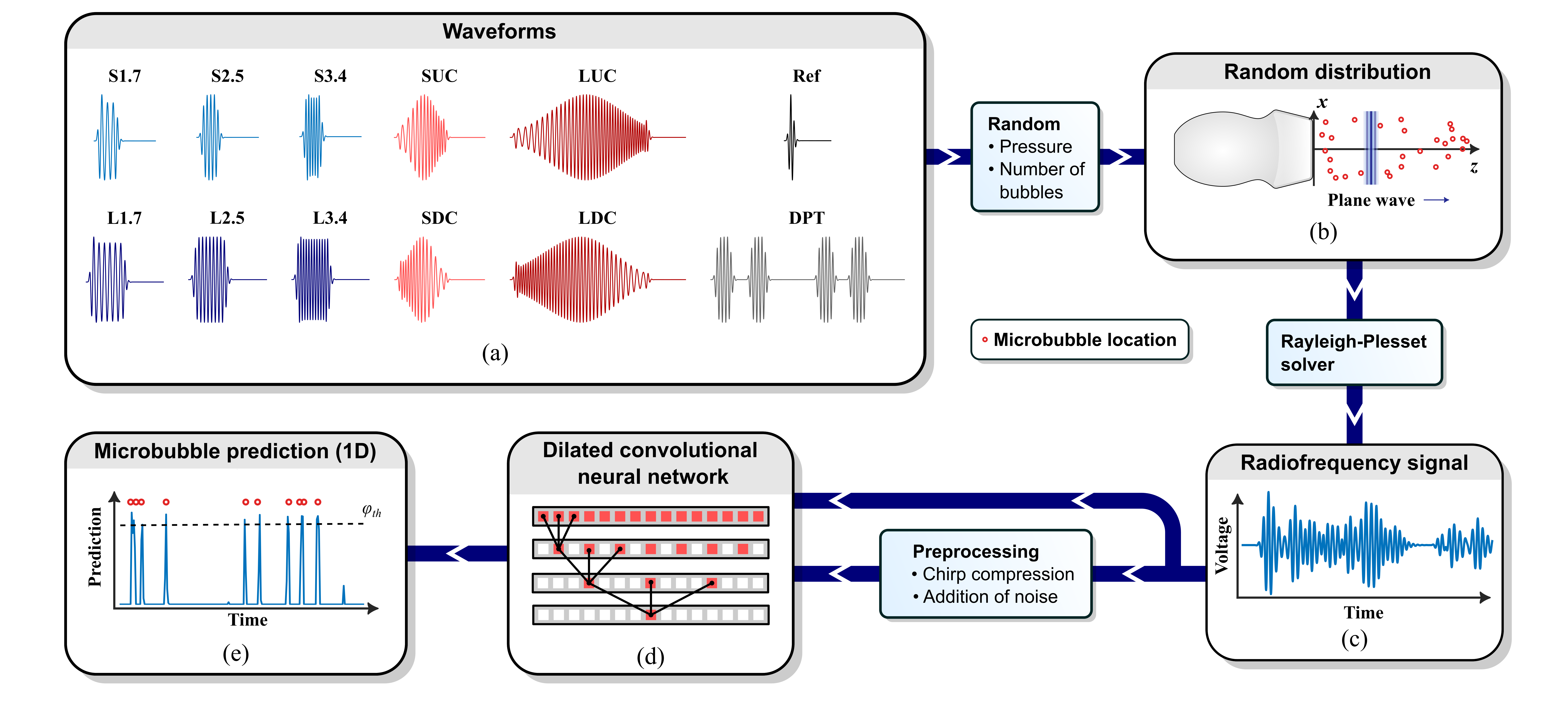}
    \caption{\textbf{Super-resolution localization pipeline.} A random distribution of microbubbles is stimulated with one of the selected pulses in (a) using a virtual P4-1 transducer (b). The simulator computes the local shapes of the pressure wave by accounting for nonlinear propagation in the medium and computes the microbubble scatter by solving the Rayleigh-Plesset-equation. The received RF signal by the transducer (c) is used as input for the dilated-CNN (d), which consequently outputs a 1D microbubble prediction (e). The network is fed by unique RF signals during training, validation, and testing.}
    \label{fig:methods}
\end{figure*}

To create super-localization CNNs for different transmit waveforms, we first define waveforms with a specific interest in ultrasound imaging.
Subsequently, we simulate RF signals generated in response to these transmit waveforms.
We then train CNN models to locate the microbubbles in the simulated RF data, where each CNN is trained on a synthetic training set corresponding to a specific waveform.
The present section explains these steps in this order.

\subsection{Waveforms}
\label{subsec: Waveforms}
We define pulses that vary in pulse length, center frequency, and rate of change of instantaneous frequency.
More specifically, we use harmonic pulses, frequency sweeps, and a pulse train.
Table~\ref{tab:pulse_list} lists the characteristics of these pulses. Fig.~\ref{fig:methods}(a) visualizes the pulses, which we divide into four categories.

The first category is a broadband imaging pulse which is the response of the transducer to a single-cycle 2.5-MHz sine wave. It is therefore a close approximation to the impulse response of the transducer.
This pulse is similar to the imaging pulse used in~\cite{Blanken2022}, and here we use it as a reference (Ref) to which the other waveforms will be compared.

The second category of pulses consists of short and long harmonic (i.e., single-frequency) pulses, with center frequencies $f_0$ of 1.7, 2.5, and 3.4~MHz.
We selected these three center frequencies for their specific physical relevance: 1.7~MHz is the resonance frequency of 2.4-µm radius microbubbles (the size used in this study), 2.5~MHz is the center frequency of the P4-1 transducer, and 3.4~MHz is twice the resonance frequency of the microbubbles. 
Driving the microbubbles at 1.7~MHz allows for recording both the fundamental and second harmonic response of the bubbles, while 3.4~MHz allows for the recording of both the fundamental and subharmonic components.
Driving the microbubbles at 2.5~MHz allows for recording the microbubble echoes at the maximum receive sensitivity of the transducer.
The short and long single-frequency pulses are the response of the transducer to sine driving signals with rectangular envelopes of approximately 1.8 and 3.5~µs, respectively.
As these different input signals are convolved with the transmit impulse response of the transducer, the center frequency of the transmit pulses is slightly shifted toward the center frequency of the transducer because of its frequency characteristics.
This shift was compensated for by iteratively adjusting the center frequency of the input signals until the center frequency of the transmit pulses matched the target frequency $f_0$, while keeping the number of sine wave cycles in the transducer input constant. Because of this compensation, the duration (full width at half maximum, or FWHM) of the transmit pulses slightly differs from the initial input signal.
The iterative adjustment stopped once the output center frequency was within 1\% of the target frequency. 

The third category comprises chirps.
Their narrow autocorrelation could facilitate bubble localization with CNNs.
Pulse duration, bandwidth, and rate of change of the instantaneous frequency all affect the non-linear microbubble response.
Therefore, we transmit both upsweep and downsweep chirps, which are known to induce different microbubble nonlinearities around resonance~\cite{Novell2008}.

The fourth category comprises a waveform consisting of a harmonic pulse train.
We use a delay-encoded pulse train (DPT) with delays optimized such that the signal-to-noise ratio (SNR) of the DPT is highest in B-mode imaging \cite{Nawijn2024}.

\begin{table}[t!]
    \centering
    \caption{List of pulses and their characteristics.}
    \begin{tabular}{llcccl}
    \\\toprule
    \textbf{Identifier} & \textbf{Waveform type} & \textbf{$f_0$ (MHz)} & \textbf{$t_\mathrm{FWHM}$ (µs)}\\\midrule
    Ref & Reference pulse & 2.5 & 0.6 \\
    \midrule
    S1.7 & Short single-frequency & 1.7 & 2.1 \\
    S2.5 & Short single-frequency & 2.5 & 1.8 \\
    S3.4 & Short single-frequency & 3.4 & 1.9 \\
    L1.7 & Long single-frequency & 1.7 & 3.8 \\
    L2.5 & Long single-frequency & 2.5 & 3.6 \\
    L3.4 & Long single-frequency & 3.4 & 3.7 \\
    \midrule
    SDC & Short downsweep chirp & 4.0 -- 1.2 & 3.3 \\
    LDC & Long downsweep chirp & 4.0 -- 1.2 & 9.7 \\
    SUC & Short upsweep chirp & 1.2 -- 4.0 & 3.3 \\
    LUC & Long upsweep chirp & 1.2 -- 4.0 & 9.7 \\
    \midrule
    DPT & Delay-encoded pulse train & 2.5 & 17.5 \\
    &(4$\times$ $\mathrm{S2.5}$)&& \\\bottomrule
    \end{tabular}\\
    $f_0$ represents the pulse center frequency, $t_\mathrm{FWHM}$ represents the pulse duration in full width at half maximum (FWHM).
    \label{tab:pulse_list}
\end{table}

\subsection{RF signal simulator}
To generate the RF signals, we use an RF signal simulator that computes the radial microbubble response by solving the Rayleigh-Plesset equation. The simulator is an extension of the simulator presented in~\cite{Blanken2022}. To increase simulation speed, we have parallelized the Rayleigh-Plesset solver. Furthermore, we have implemented a module defining the ultrasound waveforms, including the optimization algorithm described in Section \ref{subsec: Waveforms}.

In each simulation, the simulator propagates one of the defined pulses with a random acoustic pressure between 5 and 250~kPa through a random microbubble suspension comprising 10 to 1000 bubbles. The bubbles are monodisperse (mean radius of 2.4 $\mu$m) and their individual initial radius is randomly drawn from the Gaussian distribution (with a standard deviation of 5\%).
The bubbles are randomly distributed throughout the domain and the scattered pressure $P_\mathrm{s}$ is computed for each microbubble by solving the Rayleigh-Plesset equation using an ordinary differential equation solver.
The RF signals captured by a single element of the virtual P4-1 transducer are then constructed by summing the signals scattered by the amplitude and phase response of each bubble.
The imaging depth is set to 100~mm.
The microbubbles are seeded within a depth of 3.7 to 85~mm. The prior boundary ensures that the bubble signal is not extremely large due to its 1/depth dependency, the latter ensures that none of the microbubble signatures are truncated.
For each random bubble distribution, RF lines are simulated for each of the pulses such that the network training, as described in Section~\ref{subsubsec: training}, is performed on the same microbubble distributions.

\subsection{Preprocessing}
\subsubsection{Chirp decoding}
Coded excitation pulses are typically long, resulting in a low resolution if used unprocessed for delay-and-sum (DAS) image reconstruction. 
Decoding steps in conventional chirp imaging, however, improve axial resolution and lower the sidelobe level~\cite{Sun2007}. We might therefore expect that incorporating this decoding process in our pipeline facilitates the localization task of the neural network.
To evaluate the performance of models trained on chirps objectively and to understand the importance of this decoding step, we create a matched filter to decode the chirps.
This matched filter is constructed by convolving the transmitted pulses with the receive transfer function of the transducer and applying time reversal.
The RF signals are decoded by convolving them with the matched filter, such that we have a dataset with and without decoded chirps. 

\subsubsection{Noise}
\label{subsubsec: Noise}
To study the impact of noise on the performance of our super-localization pipeline, we create a series of models, each trained on a specific pulse and specific noise level.
Random normal (Gaussian) distributed noise is added to each RF signal at each epoch as a way of data augmentation to mimic the noise generated by the receiving electronics.
The standard deviation of the noise is specified as a percentage of a reference value $U_\mathrm{ref}$.
$U_\mathrm{ref}$ is defined as the signal amplitude received from a bubble in the middle of the domain (5~cm), driven by a reference pulse (Ref) with an amplitude of 125~kPa.
Before training, the noisy RF signals are filtered using a 4th-order low-pass Butterworth filter to mimic the bandpass filter that is applied in typical ultrasound scanners, including our Verasonics Vantage machine.

\subsection{CNN-based localization}
\label{subsec: AI}
Like in~\cite{Blanken2022}, we only use the RF signals received by the central element of the transducer to avoid using redundant or correlated information during training.
We utilize a CNN architecture consisting of 1D dilated convolutional layers with increasing dilation rates.
The dilated architecture allows for a receptive field of 8191 samples while keeping the network size limited (only 12 convolutional layers).
The network uses batch normalization and ReLU activation functions.
Furthermore, signals are preprocessed using 1D reflection padding to ensure that the output signal length matches the input signal length.
The output of the neural network is a prediction of a microbubble presence in 1D, expressed as one single value for each point in the RF signal. 
This output is generated within a few milliseconds, making the approach compatible with real-time imaging.

\subsubsection{Training}
\label{subsubsec: training}
For each waveform type listed in Table~\ref{tab:pulse_list}, we train a new model to deconvolve RF signals.
Besides training a network on raw RF signals obtained by transmitting chirps, we also train a network on the decoded chirp signals.
The simulated RF lines are divided into a training, a validation, and a test dataset.
The training dataset consists of 1024 samples and both the validation and test dataset consist of 960 samples.

Microbubble location prediction can be viewed as a classification task, where the ideal output is a binary array indicating the presence of a microbubble for each grid point. This binary classification, however, is imbalanced, as the number of negatives heavily outweighs the number of positives. The Dice loss (DL) is a commonly used loss function for imbalanced classification tasks, such as medical image segmentation~\cite{Sudre2017}. Alternatively, microbubble location prediction can be viewed as a regression task: when the output of the network and the labels are convolved with a soft convolution kernel, they turn into continuous microbubble density distributions. The $L_1$ loss is a suitable soft loss function for a sparse regression.
In~\cite{Blanken2022}, the binary representation is referred to as \textit{hard labels} and the continuous representation is referred to as \textit{soft labels}. Hard labels enforce a stringent localization accuracy on the output, while soft labels enforce a high recall. To leverage the advantages of both hard and soft labels, a dual-loss function was proposed in~\cite{Blanken2022}:\\
\begin{equation}
    L = \epsilon_{1} L_{1} + \epsilon_{2} \mathrm{DL},
    \label{eq:Loss}
\end{equation} 
The fastest convergence of the dual-loss function was obtained with $\epsilon_{1} = 1$ and $\epsilon_{2} = 1.6$. It was also found that the best performance metrics were obtained with a soft convolution kernel $G = e^{aj^2}$ with $j$ the array index and $a = 0.1$.
The current study uses~(\ref{eq:Loss}) with the same values for $\epsilon_{1}$, $\epsilon_{2}$, and $a$ as~\cite{Blanken2022}. The full definition of the Dice loss, $L_1$ loss, and dual-loss components are given in~\cite{Blanken2022}.

Training is performed for 1250 epochs with an Adam optimizer and a learning rate of 0.01.
The learning rate is reduced by a factor of 10 in the last 250 epochs using a learning rate scheduler (StepLR) to stabilize the training.
Network training is performed on an NVIDIA Quadro RTX 6000 GPU and takes about one hour per network. The training curves can be found in Section A of the Supplementary Information.

\subsection{Evaluation}
\subsubsection{Quantitative}
We evaluate the performance of the models with the $F_1$ score.
The $F_1$ score is the harmonic mean of the precision and the recall:
\begin{equation}
    F_1 = \mathrm{\frac{2 \times Precision \times Recall}{Precision + Recall}}
\end{equation}
Precision and recall are computed with the true positives (TP), false positives (FP), and false negatives (FN):\\
\begin{equation}
    \mathrm{Precision} = \mathrm{\frac{TP}{TP+FP}},\quad \mathrm{Recall} = \mathrm{\frac{TP}{TP+FN}}
\end{equation} 
To determine the TP, FP, and FN counts, we apply a detection threshold $\phi_\mathrm{th}$ on the model output to classify a model prediction sample as positive, see Fig.~\ref{fig:methods}(e). We optimize $\phi_\mathrm{th}$ for every model individually on the validation dataset to maximize the $F_1$ score.

\begin{figure*}
    \centering
    \includegraphics{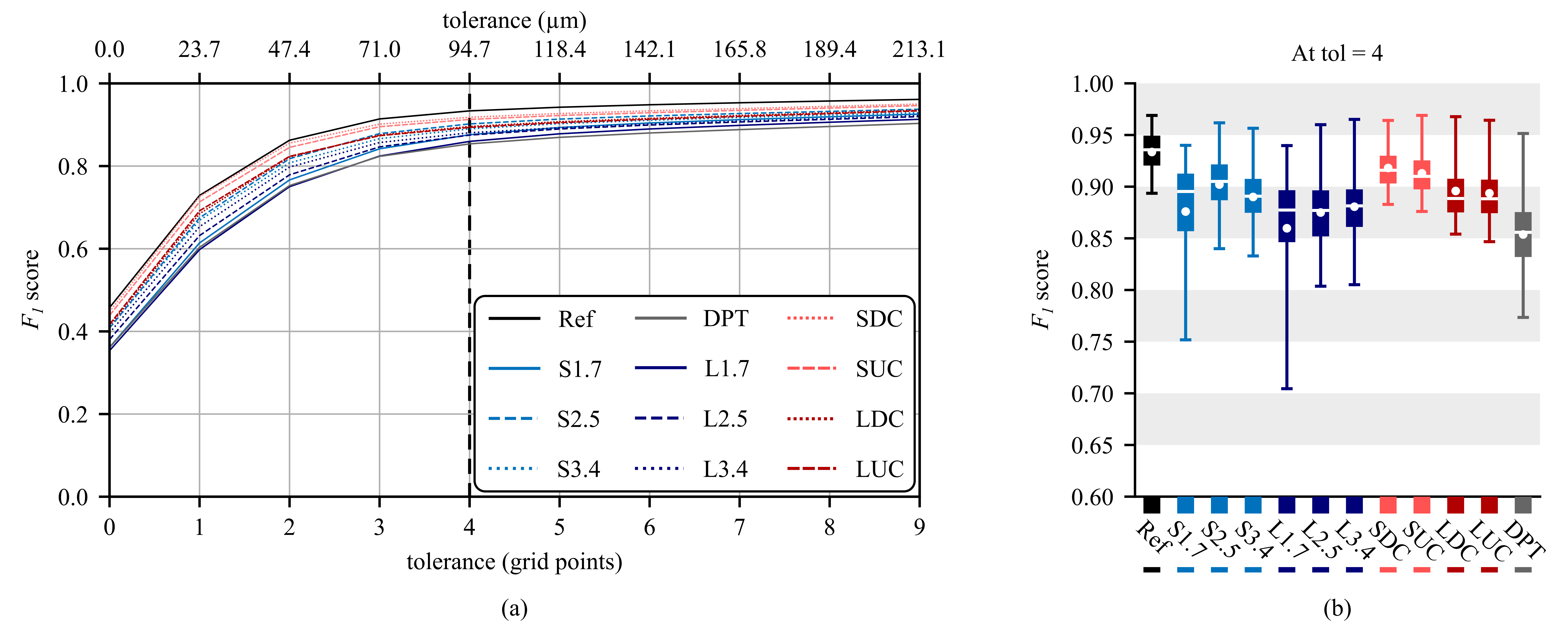}
    \caption{\textbf{Localization performance for the different models.} Performance is expressed by the $F_1$ score.
    (a) Mean performance of different models as a function of tolerance. Tolerance is expressed in both grid points of the RF signal, as well as in micrometers.
    (b) Box plot of the performance scores at a tolerance of 4 grid points for all different models. Boxes represent 25\% to 75\% of the data, whiskers represent 5\% to 95\% of the data, and white dots represent the mean of the data.}
    \label{fig:results1}
\end{figure*}

The resolution of our super-localization strategy is limited by the sampling frequency of the receiver (62.5~MHz) and the speed of sound in the medium (1480~m/s).
The highest resolution that can be obtained is thus 23.7~µm, which is smaller than the wavelength of the transmitted pulses.
Specifically, this highest resolution is about 37 times smaller than the wavelength at a transmit frequency of 1.7~MHz and 18 times smaller at 3.4~MHz.
Detection efficiency, quantified through the $F_1$ score, and resolution are linked through the tolerance allowed in quantifying true positives: a tighter requirement on location accuracy results in a lower number of true positives.
We therefore consider the tolerance, i.e., the level of super-resolution required from the neural network, as one of the parameters to investigate. This tolerance is expressed in a number of grid (or time) points in the received RF signal \cite{Blanken2022}. Ultimately, this tolerance may very well be chosen depending on the requirement of the specific application or measurement, i.e., depending on whether resolution or coverage matters most.

\begin{figure}
    \centering
    \includegraphics{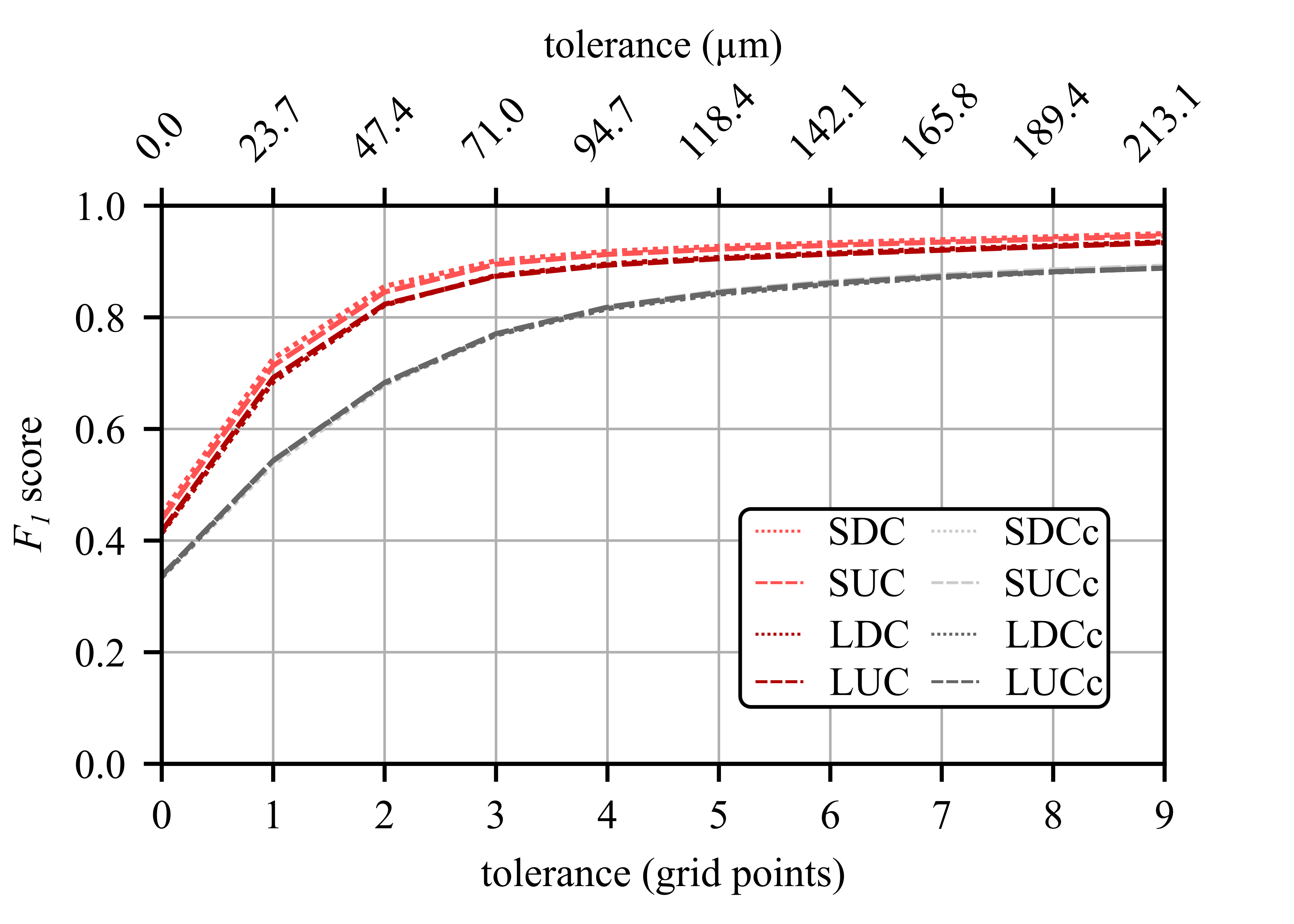}
    \caption{\textbf{Influence of chirp decoding on localization performance}. Red and gray models represent the performance of models trained on raw and decoded RF signals, respectively.}
    \label{fig:results2}
\end{figure}

\subsubsection{Qualitative}
The $F_1$ score provides an evaluation metric on single RF lines.
However, it only provides an indirect quality assessment with respect to the final images.
We therefore also qualitatively assess the performance of the algorithm on the reconstructed image by generating 2D RF signals using our simulator. We apply the trained model to all single-element RF lines to obtain a 2D microbubble prediction. Consequently, we choose the tolerance criterion and set all negative predictions to zero, e.g., the predictions below the optimized threshold criterion for this tolerance. The remaining 2D predictions are used to reconstruct super-resolved images using DAS beamforming.
As the pixel intensities may vary depending on the simulation parameters, we develop an adaptive colormap based on the statistical distribution of the pixel values in the image.
Assuming a Gaussian distribution (see Section~B in Supplementary Information), we define the lower and upper limit of the colormap as $\mu + \sigma$ and $\mu + 8\sigma$, respectively, where $\mu$ is the mean of the intensity distribution and $\sigma$ is the standard deviation.
This definition selects the bubbles predicted with sufficient confidence above the background prediction noise.
Pixel intensities lower than 0.001 are discarded in determining $\mu$ and $\sigma$ as they correspond to the large number of pixels where no bubble is predicted (at least for sparse microbubble populations).

\begin{figure*}
    \centering
    \includegraphics[width=1\linewidth]{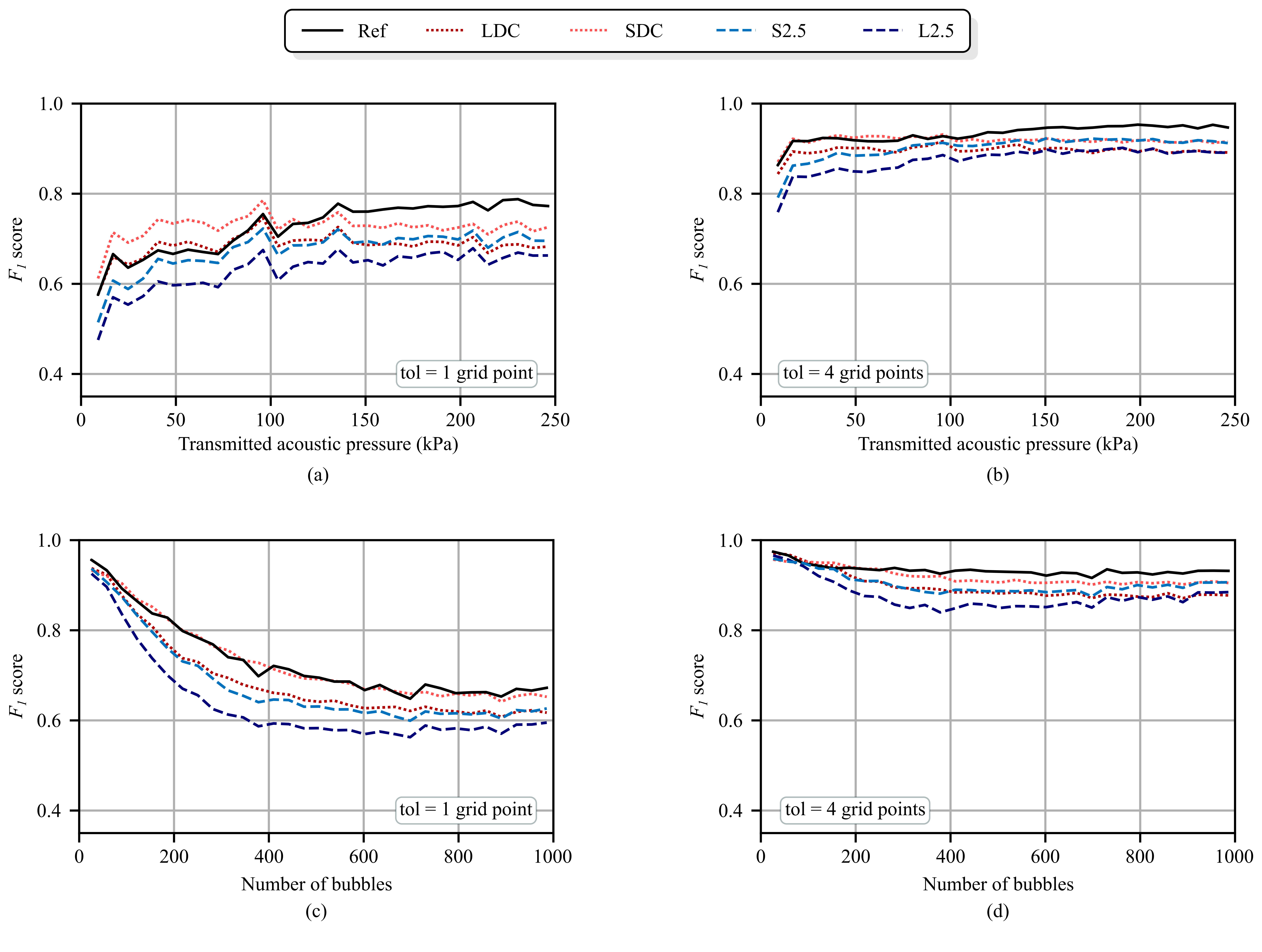}
    \caption{\textbf{Localization performance of the model as a function of the simulation characteristics for a tolerance of 1 and 4 grid points.} (a) and (b) display the relationship between transmitted acoustic pressure and model performance. (c) and (d) Represent the relationship between bubble concentration and model performance.} 
    \label{fig:Results_NB_pressure}
\end{figure*}

\subsection{Experimental validation}
To obtain a preliminary assessment of how our current approach deals with real-world complexities, we apply it to experimental data acquired when stirring a monodisperse microbubble solution (mean bubble radius = 2.8~µm, standard deviation = 0.2~µm) in a container (depth = 10~cm, width = 7.5~cm, height = 3.5~cm). A digital filter is applied to remove frequencies below 0.5~MHz from the RF signals as they fall outside the bandwidth of the P4-1 transducer and can thus be considered artifacts or noise. Also, we define a region of interest from 1.2~cm to 7.1~cm in depth to reduce reflection artifacts corresponding to the wall of the container and the magnetic stirrer. We train networks on the corresponding microbubble size distribution, transmit waveforms, and level of noise acquired in receive-only measurements.

\section{Results}

\subsection{Pulse-dependent model performance}
Fig.~\ref{fig:results1}(a) displays the trade-off between resolution and microbubble localization performance for the models.
All models show the same convergence behavior, and thus all models learn to locate the microbubbles.
Yet, the models trained on the reference pulse have the best localization performance, and they are closely followed by the models trained to deconvolve short chirps.
This, in itself, is significant, since the short chirps are 5.5 times longer than the reference pulse with 3.3~µs and 0.6~µs, respectively.
This also indicates that coded pulses, that feature narrow autocorrelations, do facilitate CNN-based deconvolution.
Additionally, a similar performance can be seen for networks trained on upsweep and downsweep chirps.

These observations are further confirmed in Fig.~\ref{fig:results1}(b), which compares the statistical distributions of the performance across test samples for all models and a tolerance of 4 grid points.
The pulses with the highest mean performance are also those with the least sample-to-sample variation.
In all cases, the lower the mean performance of the model, the larger the variability of the prediction accuracy over the test samples. Interestingly, the models trained on 1.7~MHz short pulses show lower performance (i.e., mean and deviation) compared to models trained on the reference pulse, but also compared to models trained on the other short pulses with center frequencies of 2.5 and 3.4~MHz.
Models trained on the pulse train offer a low localization performance, which suggests that the deconvolution task is too complex to exploit the potential of pulse trains, at least with the current network architecture.

Fig.~\ref{fig:results1} displays the deconvolution results on raw RF signals only, which implies that the tasks performed by the network encompass the complete deconvolution process including, arguably, the steps that are needed for conventional image reconstruction.
For example, conventional chirp imaging must utilize chirp decoding algorithms to recover the imaging resolution.
To investigate whether performing this compression prior to deep learning-based deconvolution improves localization performance, we applied a chirp decoding filter as described in Section \ref{sec: Methods} to decode the chirp training dataset and train a network on these alternative, decoded RF signals, as described in Section~\ref{subsubsec: training}. Fig.~\ref{fig:results2} compares the performance of the models trained on raw RF lines with that of the models trained on decoded RF lines. The latter are denoted with a 'c' behind the identifier. As can be seen in the plot, the performance of all models trained on decoded RF signals is similar and much lower than that of the models trained on raw RF signals. We expect that the origin of this performance drop is twofold. On the one hand, imperfect decoding of the RF signal, i.e., due to either resonant microbubble scattering, ultrasound attenuation, or nonlinear propagation, results in small-amplitude but long-range artifacts. The frequency content of these artifacts is not random, and they are thus difficult to distinguish from the signature of a weak scatterer. On the other hand, convolving the RF lines with the decoding filter may result in the loss or delocalization of features that are characteristic of bubble contrast, such as the nonlinearities that give rise to the generation of harmonics. We will elaborate on these aspects in the Discussion. In fine, and surprisingly, decoding the chirps before deep learning-based deconvolution makes the task far more difficult for the model rather than easier.

To explore the performance of the models in more detail, we select five models: Ref, S2.5, L2.5, SDC, and LDC.
Ref, the model trained on the reference, displays the best localization performance.
S2.5 is the most promising harmonic pulse, likely because their frequency content matches that of the transducer. Additionally, we select L2.5 as it is the longer equivalent of S2.5. SDC and LDC are selected because of their high localization performance and unique time-frequency distribution. The other models either offer a similar performance or a performance that is too low to bear further consideration.

\subsection{Performance versus pressure and density}
The response of microbubbles to acoustic stimuli depends on the transmitted acoustic pressure. Increasing acoustic pressure generally results in an increased radial excursion, enhancing backscatter and visibility of microbubbles in the RF lines. Increasing the acoustic pressure also changes the nature and strength of the bubble nonlinearities: for pressures between approximately 5 to 25~kPa the bubbles are highly non-linear because the size-dependent surface tension imparted by the shell plays a large role \cite{Versluis2020}. The nonlinearities introduced by the shell, however, become less pronounced when the pressure is further increased \cite{Overvelde2010}. From pressures ranging approximately from 50 to 100~kPa, microbubble scatter increases, with mostly a fundamental response. For pressures above 100~kPa, intrinsic bubble nonlinearities take precedence over shell effects and induce the generation of both harmonic and subharmonics, making the resonance amplitude pressure-dependent and skew the resonance curve \cite{Versluis2020}. The combination of these effects makes the bubble signatures highly specific and, presumably, easier for the network to deconvolve. Fig.~\ref{fig:Results_NB_pressure} (a) and (b) display the deconvolution performance as a function of the transmitted acoustic pressure on the test dataset for a localization tolerance of 1 and 4 grid points, respectively. For pressures ranging from 25 to 100~kPa, i.e., in the range for quasi-linear bubble response, the performance of all networks is slightly poorer than for higher pressures. The difference, however, is mild. Note that pressures below 25~kPa are hardly used for imaging in practice, as they provide insufficient SNR. The performance trend for increasing bubble density and increasing pressure is visible for all models, as demonstrated in Section C of the Supplementary Information.

\begin{figure*}
    \centering
    \includegraphics[width=1\linewidth]{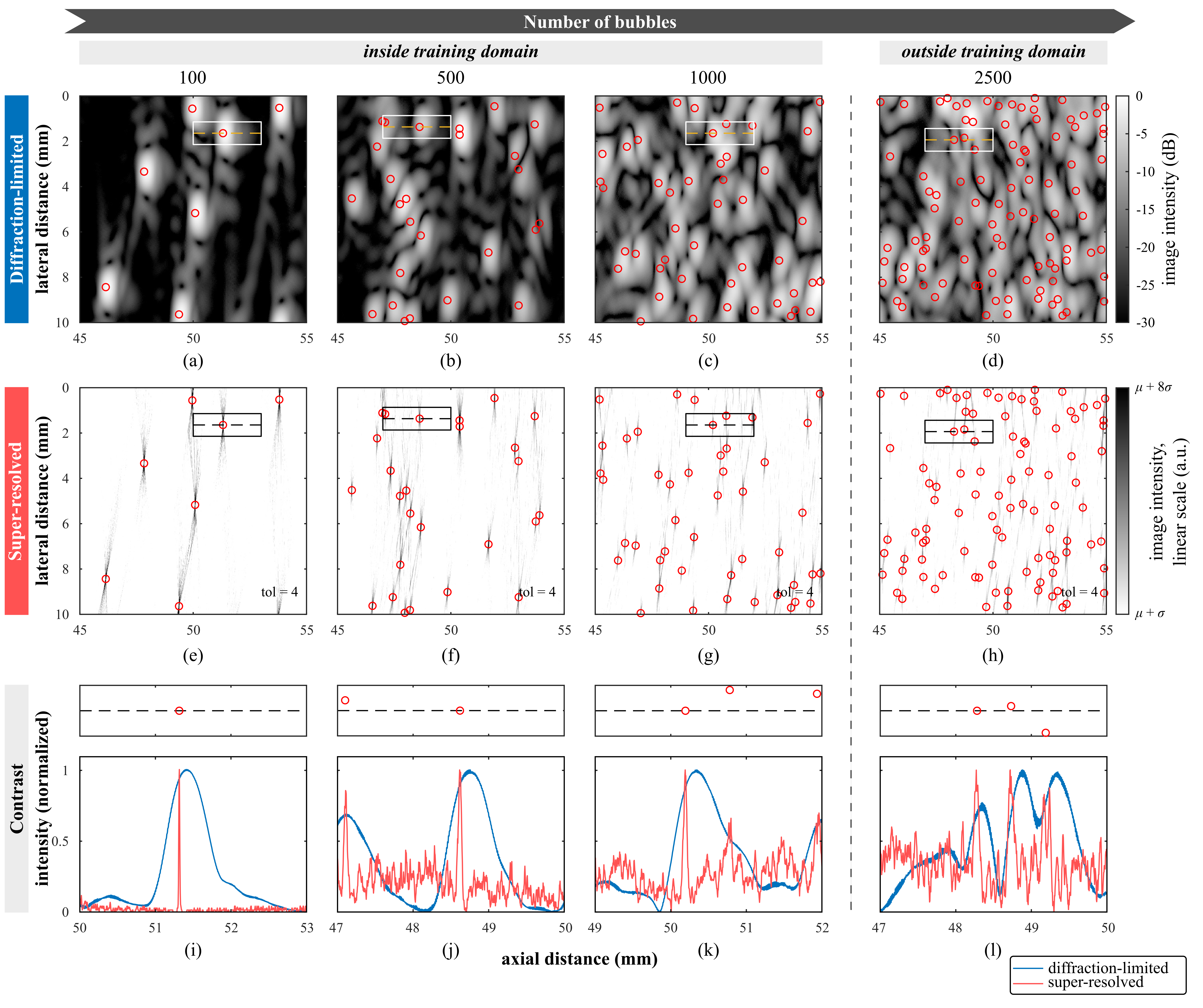}
    \caption{\textbf{Images of diffraction-limited and super-resolved beamforming for an increasing bubble density.} Images are generated using delay-and-sum of the raw RF lines (diffraction-limited, (a) to (d)) and responses of a model trained on the Ref pulse (super-resolved, (e) to (h)). Both the diffraction-limited and the super-resolved image are a segment in the middle of the full image domain. The columns represent the number of bubbles in the domain (100, 500, 1000 and 2500). The contrast in the super-resolved image is scaled to the mean ($\mu$) and standard deviation ($\sigma$) of the pixel intensities larger or equal than 0.001 and the color bar is inverted for better visualization. The red circles represent the ground truth bubble location. The bottom row plots the scaled image intensity of the cross-sections in the diffraction-limited image (blue) and the super-resolved image (red).}
    \label{fig:nBubblesComparison}
\end{figure*}

The length of the various pulses ranges from 37 samples (Ref) to 1091 samples (DPT). Due to the inertial and resonant effects of microbubble oscillations, the microbubble response itself can be even longer. As the bubble density increases, the bubble echoes will increasingly overlap in the RF signals. This high signal density and the associated interference between bubble responses increase the complexity of the task assigned to the neural networks. Note that sparsity in the bubble echoes is, arguably, the most stringent requirement of more traditional approaches to ultrasound super-localization. Fig.~\ref{fig:Results_NB_pressure}(c) and (d) present the model performance versus the number of microbubbles in the simulation. For a tolerance of 1 grid point, the performance for all pulses sharply decreases until about 400 bubbles per RF signal before stabilizing to an $F_1$ score of about 0.6. Thus, for lower tolerances, the localization performance is more affected by bubble density than by the choice of the ultrasound pulse or ultrasound pressure. 
The same general trend is observed for a tolerance of 4 grid points. However, the $F_1$ score is far less affected and retains a value of approximately 0.9, even for the highest bubble concentration. On the contrary to the 1-grid-point-tolerance case, the choice of transmit pulse has more importance than the number of bubbles. In all cases, the model trained on long harmonic pulses is the most affected by both acoustic pressure and the number of bubbles, while the model trained on short reference pulses gives the most consistent results. The other pulses display intermediate performance, with the important remark that the performance of the model trained on short chirps is very close to (and occasionally exceeds) that of the reference pulse.

\begin{table}[h]
    \centering
    \caption{Mean and standard deviation in nonzero pixel intensity in Fig.~\ref{fig:nBubblesComparison}.}
    \begin{tabular}{lccccl}
    
    & \multicolumn{4}{c}{number of bubbles}\\\toprule
    \textbf{} & \textbf{100} & \textbf{500} & \textbf{1000} & \textbf{2500}\\\midrule
    $\mu$ & 1.899 & 6.817 & 11.425 & 15.761 \\
    $\sigma$ & 1.487 & 3.216 & 4.005 & 4.559\\
    $\sigma/ \mu$ & 0.783 & 0.472 & 0.351 & 0.289\\
    \bottomrule\\
    \end{tabular}\\
    Pixel intensity distribution of images containing a varying number of bubbles for the Ref model. $\mu$ represents the mean intensity, $\sigma$ represents the standard deviation.
    \label{tab:int_list1}
\end{table}

\begin{figure*}
    \centering\includegraphics[width=1\linewidth]{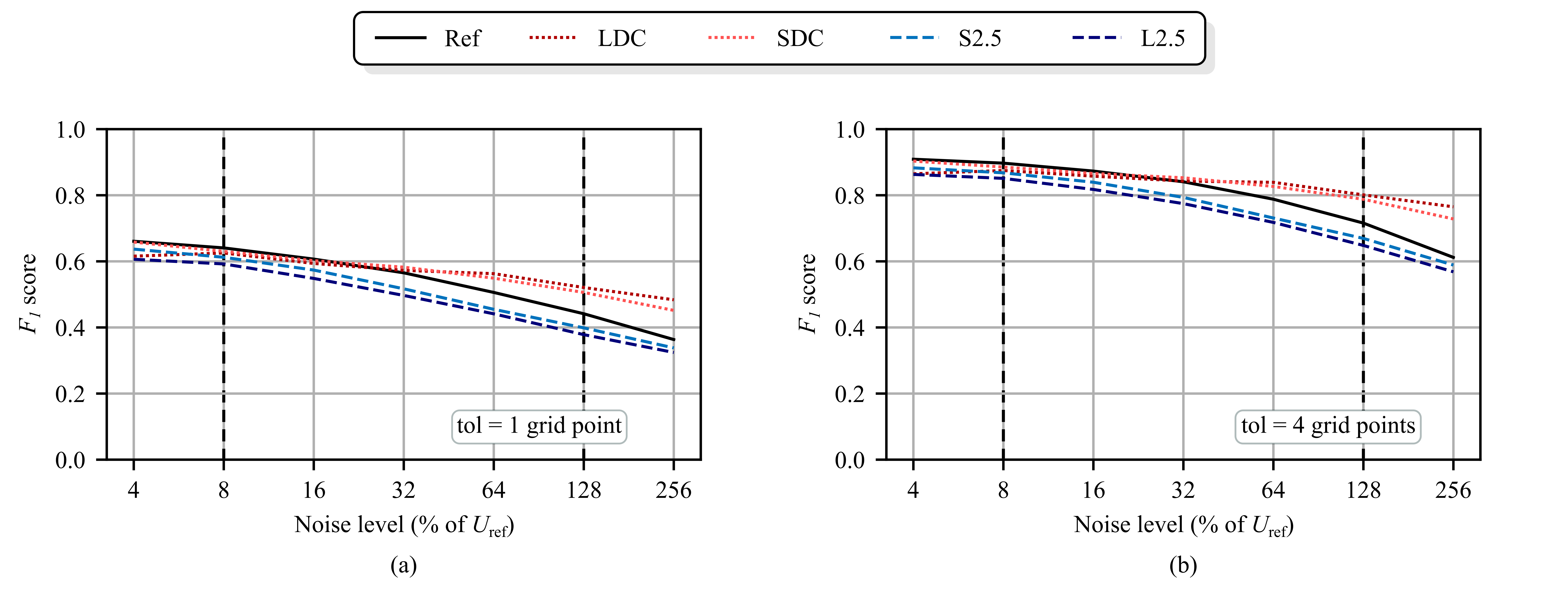}
    \caption{\textbf{Localization performance of the models trained on a selection of pulses for exponentially increasing noise levels (4\% to 256\% of the reference voltage $U_\mathrm{ref}$).} (a) Performance at a localization tolerance of 1 grid point. (b) Performance at a localization tolerance of 4 grid points.}
    \label{fig:NoiseEvaluation}
\end{figure*}

\begin{figure}
        \centering\includegraphics[width=1\linewidth]{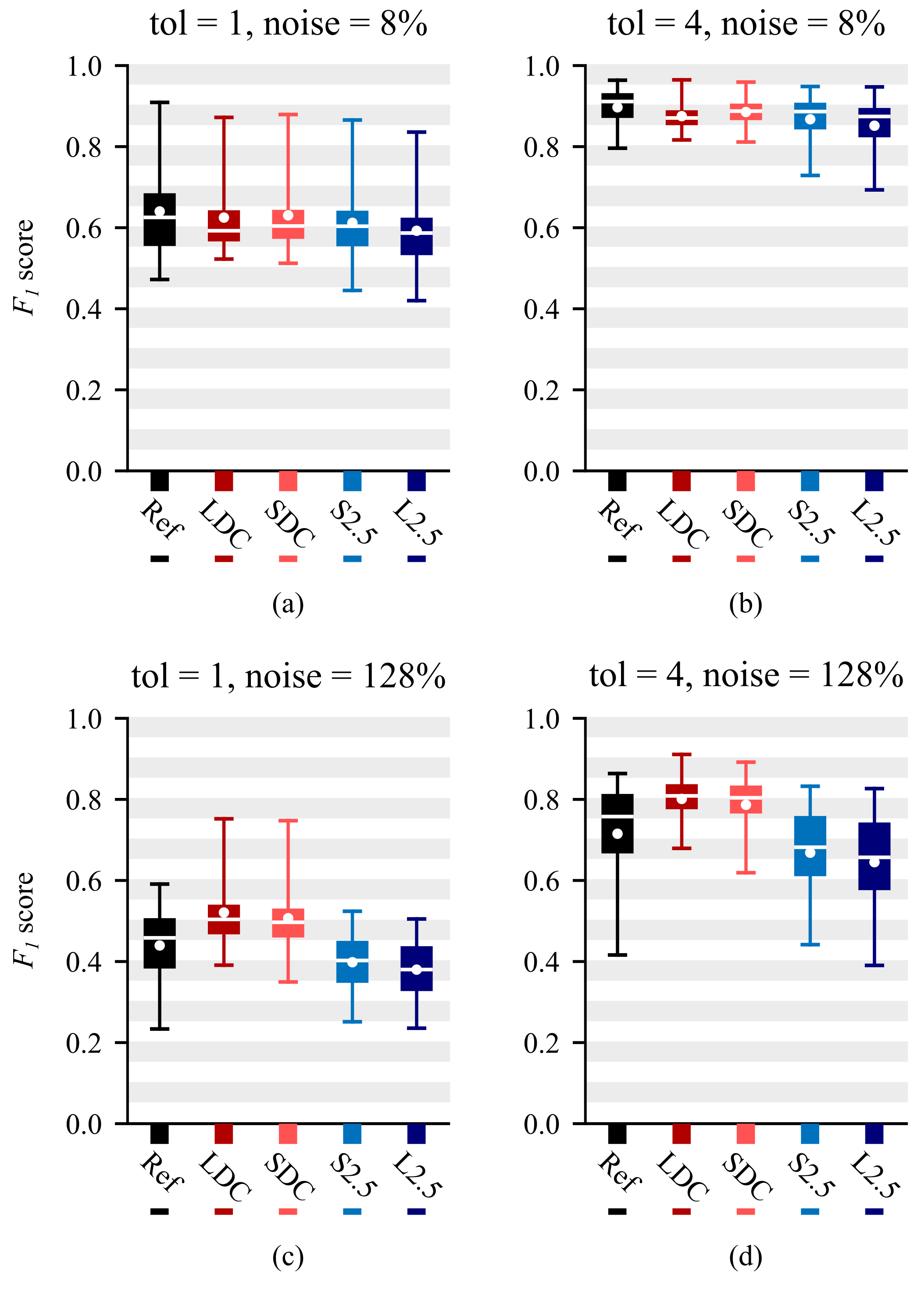}
        \caption{\textbf{Performance distributions of the selected models from Fig.~\ref{fig:NoiseEvaluation} on the test dataset for two tolerances at two noise levels. } (a) tolerance = 1, noise = 8\% of $U_{\mathrm{ref}}$, (b) tolerance = 4, noise = 8\% of $U_{\mathrm{ref}}$, (c) tolerance = 1, noise = 128\% of $U_{\mathrm{ref}}$, (d) tolerance = 4, noise = 128\% of $U_{\mathrm{ref}}$.}
        \label{fig:NoisePerformanceDetailed}    
\end{figure}

The $F_1$ score is a valuable metric to quantify the performance of the networks that operate on single RF signals, but it can not be directly interpreted in terms of ultrasound image quality. To that end, Fig.~\ref{fig:nBubblesComparison} shows both diffraction-limited and super-resolved images reconstructed using raw and deconvolved RF lines generated in response to the short imaging pulse (reference pulse). Fig.~\ref{fig:nBubblesComparison}(a) to (c) on the one hand, and Fig.~\ref{fig:nBubblesComparison}(e) to (g) on the other hand, correspond to increasing number of bubbles (100 to 1000) within the range of bubble densities represented in the training set. Fig.~\ref{fig:nBubblesComparison}(d) and (h) show an example of generalization, i.e., a number of bubbles far beyond the training range (2500 bubbles). The ground truth bubble locations in Fig.~\ref{fig:nBubblesComparison}(a) to (h) are denoted by the empty red circles. Deep learning-based deconvolution thus also provides accurate super-resolved images, based on a qualitative assessment of the reconstructed images. Table \ref{tab:int_list1} lists the mean and standard deviation of the nonzero pixel intensities. The ratio of $\sigma$ and $\mu$ decreases with the number of bubbles in the image, indicating that the contrast between the bubbles and the background decreases. Histograms of the pixel intensity distributions can be found in Fig. S2 of the supplementary information.

The intensity of the cross-section depicted by the dashed lines in Fig.~\ref{fig:nBubblesComparison}(a) to (h) are displayed in Fig.~\ref{fig:nBubblesComparison}(i) to (l). The deep learning models greatly improve the axial resolution as compared to diffraction-limited imaging. Although the background prediction noise increases with increasing microbubble density, the presence of the bubbles is predicted with excellent contrast even outside the training range. 
Increasing the bubble density also leads to a slight loss of contrast precision, visible in the FWHM of the point spread functions.
The FWHM becomes gradually wider, measuring about 15~µm in Fig.~\ref{fig:nBubblesComparison}(i) (100 bubbles), approximately 50~µm in Fig.~\ref{fig:nBubblesComparison}(j) (500 bubbles) and Fig.~\ref{fig:nBubblesComparison}(k) (1000 bubbles), and 90~µm in Fig.~\ref{fig:nBubblesComparison}(l) (2500 bubbles).

\begin{figure*}[ht]
    \centering
    \includegraphics[width=1\linewidth]{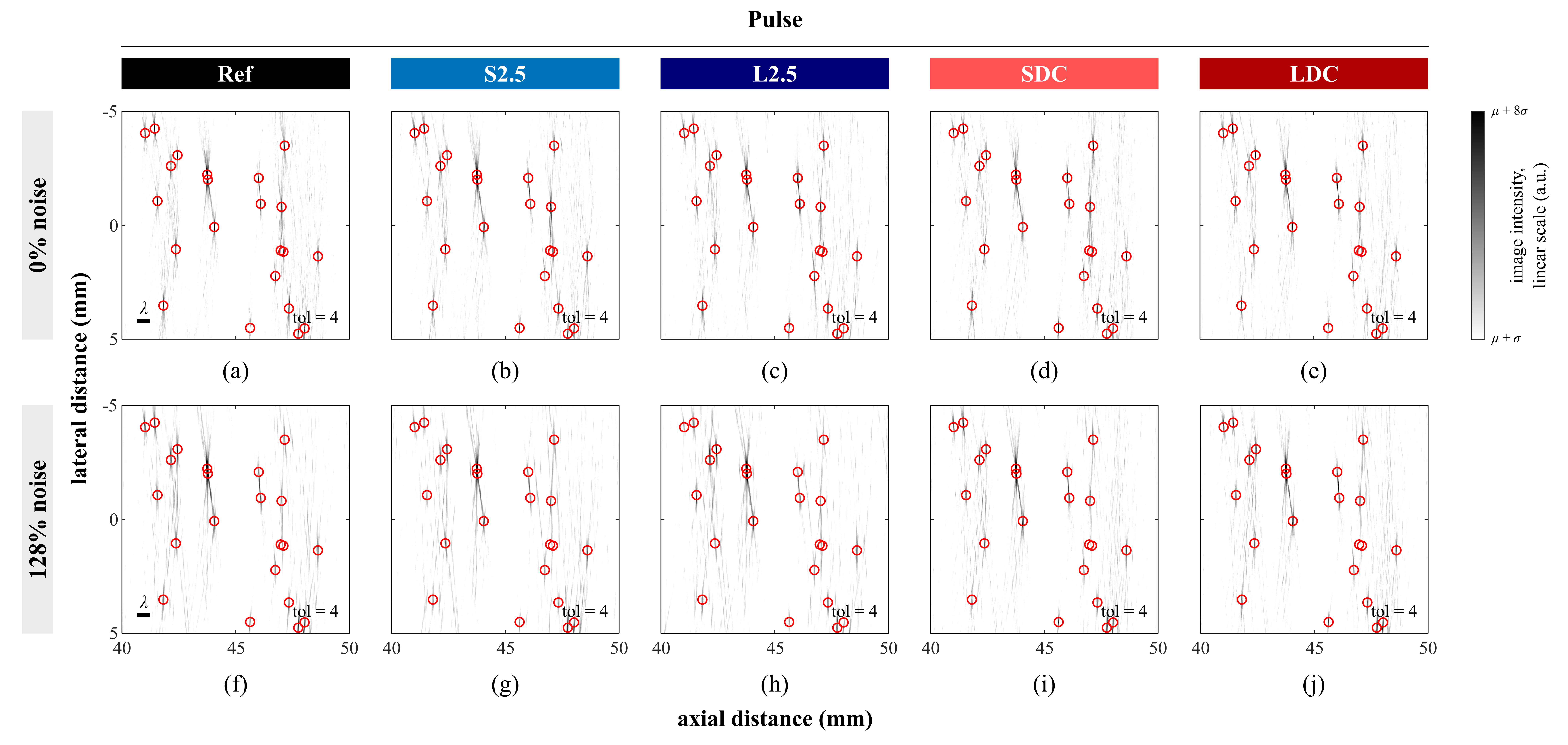}
    \caption{\textbf{Delay-and-sum generated images of midfield responses of models trained on different pulses.} The red circles represent the ground truth bubble location. The entire domain contains 500 randomly seeded bubbles and the transmitted acoustic pressure was 125 kPa. The color bar representing the image intensity is kept the same for the models that were trained to deconvolve the same pulse.}
    \label{fig:pulseComparison}
\end{figure*}

\subsection{Robustness to noise}
\label{subsec: Robustness_to_noise}
The models evaluated so far are trained and tested on noise-free RF signals. Evaluating the applicability of our super-resolution pipeline to real-world ultrasound machines requires considering noise, which is both pronounced and unavoidable in ultrasound data. Fig.~\ref{fig:NoiseEvaluation} shows the evolution of the deconvolution performance as a function of the level of noise, which is defined as random, Gaussian-distributed noise. 
As described in Section~\ref{subsubsec: Noise}, one model was trained for each pulse and each specific level of noise, so each point in the graph represents the performance of that corresponding model. 
The noise levels were 4\%, 8\%, 16\%, 32\%, 64\%, 128\%, and 256\% with respect to $U_\mathrm{ref}$ defined in Section~\ref{subsubsec: Noise}.

As expected, the model trained on the reference pulse still outperforms all other models for low noise levels (4\% and 8\%) at both a tolerance of 1 and 4 grid points. However, the models trained to deconvolve the chirps (LDC and SDC) maintain their performance up to a noise level of 64\% and only show a mild degradation thereafter. Consequently, both exceed the performance of the model trained on the reference pulse above a noise level of 32\%. S2.5 and L2.5 also rapidly lose their performance for increasing noise, and their performance remains below that of the reference pulse model for any noise level.

To visualize the practical implications of this loss of performance, Fig.~\ref{fig:pulseComparison} displays DAS images reconstructed using deconvolved, noise-free RF lines and RF lines augmented with 128\% noise. The color bar is based on the pixel statistics of the noise-free images. Fig.~\ref{fig:NoiseEvaluation}(a) to (e) show the images resulting from the noise-free data and show little difference across pulse types.

For a noise of 128\%, the image contrast drops, see Fig.~\ref{fig:pulseComparison}(f) to (j). The bar indicated by $\lambda$ in Fig.~\ref{fig:pulseComparison}(a) and (f) represent the wavelength of the reference pulse, given as a reference to represent the diffraction-limited image resolution.

Table \ref{tab:int_list2} lists the mean and standard deviation of the nonzero pixel intensity in Fig. \ref{fig:pulseComparison}.

\begin{table}[ht]
    \centering
    \caption{Mean and standard deviation in nonzero pixel intensity in Fig. \ref{fig:pulseComparison}.}
    \begin{tabular}{lccccl}
    & \multicolumn{5}{c}{model}
    \\\toprule
    \textbf{} & \textbf{Ref} & \textbf{S2.5} & \textbf{L2.5} & \textbf{SDC} & \textbf{LDC}\\\midrule
    $\mu$ noise-free & 6.817 & 6.792 & 6.897 & 6.893 & 6.433\\
    $\sigma$ noise-free& 3.216 & 3.170 & 3.171 & 3.246 & 3.083\\
    $\sigma/ \mu$ & 0.472 & 0.467 & 0.460 & 0.471  & 0.479\\
    \midrule
    $\mu$ 128\% noise & 6.156 & 6.281 & 5.791 & 5.574 & 5.857\\
    $\sigma$ 128\% noise & 2.794 & 2.703 & 2.692 & 2.864 & 2.895\\
    $\sigma/ \mu$ & 0.454 & 0.430 & 0.465 & 0.514 & 0.494\\
    \bottomrule\\
    \end{tabular}\\
    Pixel intensities of different models on excitation of the same bubble distribution. $\mu$ represents the mean intensity, $\sigma$ represents the standard deviation of the intensity.
    \label{tab:int_list2}
\end{table}

\subsection{Experimental validation}

\begin{figure*}
    \centering
    \includegraphics{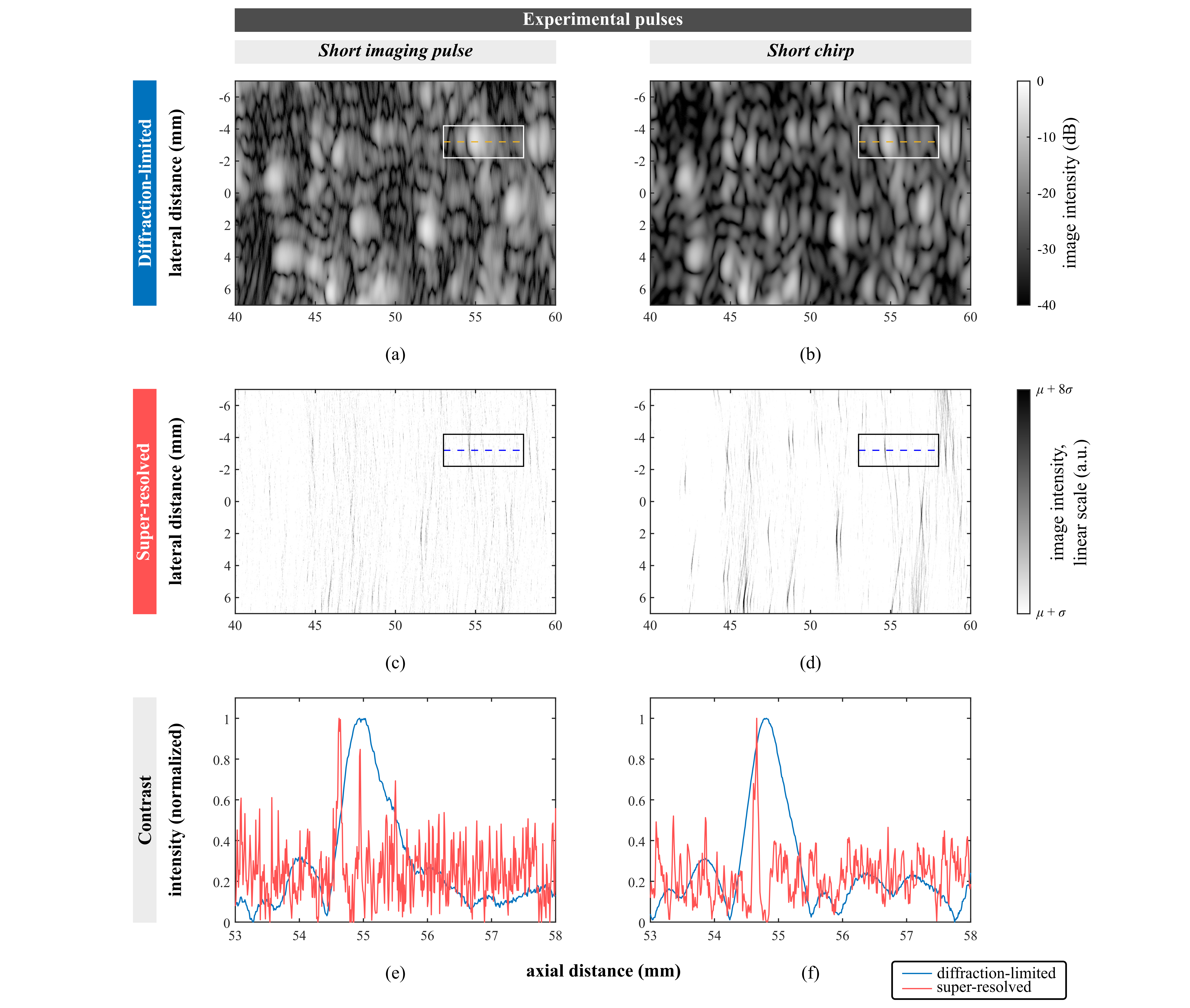}
    \caption{\textbf{Diffraction-limited ((a) and (b)) and super-resolved images ((c) and (d)) obtained in an experimental setup.}  RF lines are obtained by transmitting a short imaging pulse ((a) and (c)) and a short chirp ((b) and (d)). To recover resolution, RF lines obtained by chirps (b) are decoded before image reconstruction. (e) and (f) display the contrast of the cross-sections in the diffraction-limited and super-resolved images.}
    \label{fig:exp_reconstruction}
\end{figure*}

Fig.~\ref{fig:exp_reconstruction} displays the diffraction-limited and super-resolved images for both a short imaging pulse (SIP) (FWHM = 0.7 µs, $f_0$ = 2.5~MHz) and a short upsweep chirp (FWHM = 4.5~µs, $f_{start}$ = 1.2~MHz, $f_{end}$ = 4.0~MHz). The RF lines used for reconstructing the diffraction-limited image of the chirp are first compressed to recover optimal resolution. Fig.~\ref{fig:exp_reconstruction} shows that for both pulses, the contrast in the super-resolved images is located within the point-spread functions in the diffraction-limited images. The videos SI02\_video\_chirp.mp4 and SI03\_video\_SIP.mp4 allow to visually track the movement of the microbubbles in both the diffraction-limited and the super-resolved images. The SNR in the super-resolved chirp images is considerably higher than in the super-resolved short imaging pulse, confirming the observations in Section~\ref{subsec: Robustness_to_noise}.

\section{Discussion}

We have explored the applicability and performance of frequently used ultrasound imaging pulses for deep learning-based ultrasound super-localization. All models trained on raw RF signals learn to locate the microbubbles with high performance, i.e., an $F_1$ score of 0.85 to 0.92, at a resolution of 4 grid points. While conventional B-mode imaging often faces challenges in pulse selection due to the resolution-SNR trade-off, we have demonstrated that when using deep learning in combination with chirps, resolution, and robustness to noise can go hand-in-hand since chirps indeed outperform short imaging pulses in noisy conditions. The conventional paradigm that the resolution is equal to the pulse length thus does not hold in deconvolution-based super-resolution. While the relation between the pulse and resolution still holds, the resolution in such super-resolution imaging is ultimately governed by the performance of the deconvolution process.

\begin{figure}
    \centering
    \includegraphics{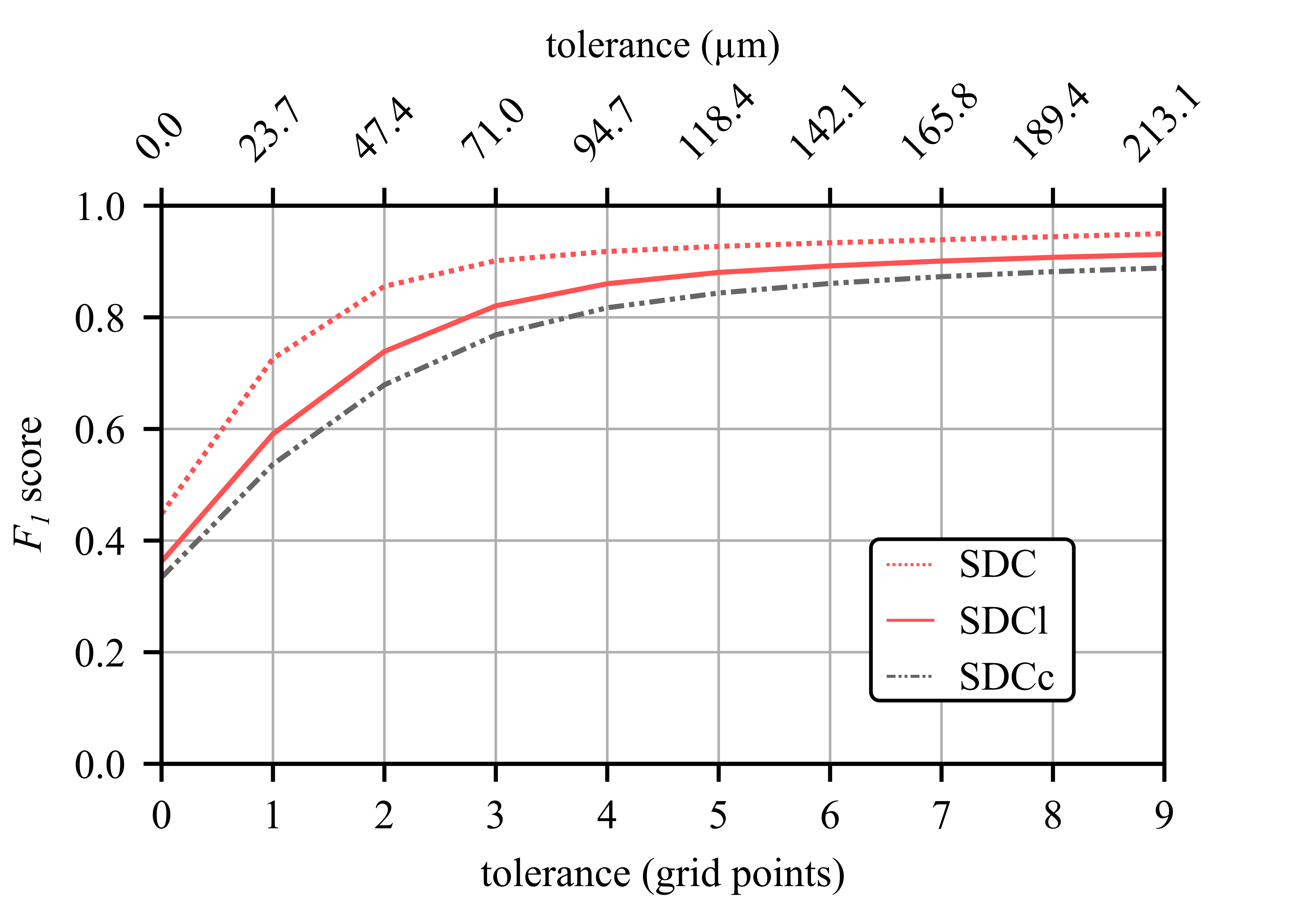}
    \caption{\textbf{Performance of models trained on transmitted SDCs.} SDC represents a model trained on raw RF signals, SDCl represents a model trained on simulated data with a linearized Rayleigh-Plesset equation, SDCc represents the model trained on decoded chirps.}
    \label{fig:SDC_comparison}
\end{figure}

Additionally, we have demonstrated the superior performance of \textit{direct deconvolution} compared to \textit{pulse decoding} prior to deconvolution when using chirped excitation. 
The linear decoding filter comprises a time-reversed version of the received pulse, as proposed by~\cite{Sun2007}, and thus depreciates the nonlinearities of the microbubble response before the signal is processed by the network.
To examine the importance of bubble nonlinearities for the deconvolution process, we have trained a model on synthetic data obtained with a linearized RP-solver. The performance of this model is shown in Fig.~\ref{fig:SDC_comparison} together with that of the models trained on the decoded but nonlinear RF signals, and that of the model trained on raw, nonlinear RF signals.
Clearly, including the bubble-specific nonlinearities improves the performance of the deconvolution task. Decoding the chirps has a much smaller effect on the performance of this task.

In Section~\ref{subsec: Robustness_to_noise}, we evaluated the robustness of our super-resolution pipeline to noise. To further assess this robustness, models can be either trained on single noise levels, or on a range of noise levels and evaluated for varying noise levels. We have tested both scenarios and discuss the performance in Supplementary Information Section E. In the first case, as shown in Fig. S5, the performance of the networks remains stable up to the noise level on which they were trained and drops after that. In the second case, as shown in Fig. S6, the differences in performance are marginal compared to the first scenario. The more concise representation presented in Section~\ref{subsec: Robustness_to_noise} achieves the highest performance in terms of $F_1$ score and therefore well represents the robustness to noise quantified in a more extensive assessment frame.

It is important to note that our approach may contain an implicit bias, since the coefficients $\epsilon_1$ and $\epsilon_2$, affecting the ratio between $L_1$ and Dice loss, are kept the same as in~\cite{Blanken2022}. These coefficients had been optimized for a short imaging pulse. However, the exact choice of these parameters was not crucial to the training performance and therefore we do not expect an important effect here as well. The same holds for other hyperparameters including the number of training epochs, training batch size, learning rate, training set size, and network depth. The fact that the hyperparameters were optimized on a data set with a short imaging pulse in~\cite{Blanken2022} may explain why the short imaging pulse (Ref) performs the best in low-noise conditions. A different ranking of the pulses may be expected when the hyperparameters are optimized for each waveform independently.

Recent studies in deep learning-based microbubble localization have explored different network architectures for improving the resolution of images for microvessel mapping~\cite{Luan2023}. The current study, however, localizes the microbubbles on raw RF signals prior to beamforming, thereby reducing the complexity of the network and providing a route to overcome the sparsity requirement of ULM-based techniques. The dilated CNN is advantageous for several reasons. First, the convolutional layers of the neural network allow for translational equivariance, meaning that a shift in bubble position results in a corresponding shift in the networks response. This makes the dilated CNN architecture useful for imaging in deep tissue. 
Second, the dilation provides a large receptive field with a limited number of layers, thereby reducing the number of network parameters and the training time. Additionally, as ultrasound images are often captured sequentially, adjacent RF signals also contain temporal information which may aid microbubble detection. This information, in fact, is essential to the spatiotemporal filtering in ULM. 

Although dilated CNNs have previously shown to have strong performance on sequential data~\cite{Bai2018}, other architectures like recurrent neural networks, long short-term memory, gated recurrent units, or transformers may aid super-localization in e.g., super-resolved vector flow images and very dense bubble distributions. The value of such architectures in this context, however, remains to be addressed.

In our approach, we train the networks to deconvolve RF signals in 1D, and thus localization is parallelized per element. Consequently, we disregard the correlation between adjacent elements. We expect that increasing the dimensions of the networks to 2D allows to exploit this information better and aid localization both in lateral as well as axial direction.

\subsection{Translation to real-world imaging}
In the case of our deep learning approach, we face a trade-off between the number of physical elements that we aim to investigate and the level of realism with which we do this. Because we investigate many parameters (i.e., here, twelve distinct pulses, a large acoustic pressure range, a large range of bubble densities, and a large range of noise levels), we compromise on the level of realism. While we have the capabilities to simulate a more realistic dataset~\cite{Blanken2024}, which includes 3D pressure fields, tissue background, and tissue attenuation, it is resource- and time-consuming. As a result, the computational cost of generating such data would prohibit investigating the performance of networks trained on individual waveforms. Moreover, the super-resolution pipeline presented in this study can already be used to track microbubbles with super-resolved precision in an experimental setting.

Nevertheless, for translation to in-vivo settings, significant challenges need to be addressed. The models trained in this work were trained on a fixed noise level to resemble machine noise. However, in-vivo SNR will vary per measurement and is influenced by local physiological conditions. Deep ultrasound imaging, such as arterial imaging for assessing flow characteristics, can therefore face very low SNR ratios, which are typically not accounted for in a lab setting. Recently developed strategies for generating realistic training datasets~\cite{Shin2024,Blanken2024} may allow for training super-resolution networks with improved performance in experimental data, and, ultimately, help to address the complexities involved in in-vivo ultrasound super-resolution.

\section{Conclusion}
In this work, we demonstrate the potential of CNN models in microbubble localization for a wide range of commonly used ultrasound transmit pulses. We show that these models are more sensitive to the density of bubbles than to the ultrasound pressure and that they can be generalized far beyond the range of bubble densities used for training. Although short imaging pulses offer optimal performance in ideal, noise-free conditions, chirps significantly improve robustness to noise, while offering nearly the same performance in a noise-free situation. Ultimately, these results demonstrate, first, that deep learning strategies aiming at deconvolving ultrasound data do not follow the paradigm linking stating that the resolution is equal to the pulse length and, second, that bubble nonlinearities, and thus bubble physics, are a strong asset for improving ultrasound super-resolution using microbubbles.

\section{Acknowledgments}
The authors would like to thank Mike Averkiou and Olivier Couture for valuable discussions.

\bibliographystyle{IEEEtran}
\bibliography{references.bib}

\newpage
\appendix

\renewcommand{\thefigure}{S\arabic{figure}}
\setcounter{figure}{0}
\renewcommand{\thetable}{S\Roman{table}}
\setcounter{table}{0}

\newcommand{\nBubblesComparison}{5}

\section*{Waveform-Specific Performance of Deep Learning-Based Super-Resolution for Ultrasound Contrast Imaging -- Supplementary information}

\noindent \textbf{
    Rienk Zorgdrager,
    Nathan Blanken,
    Jelmer M. Wolterink,
    Michel Versluis, and
    Guillaume Lajoinie
    }

\subsection{Training curves}

\begin{figure}[ht]
    \centering
    \includegraphics{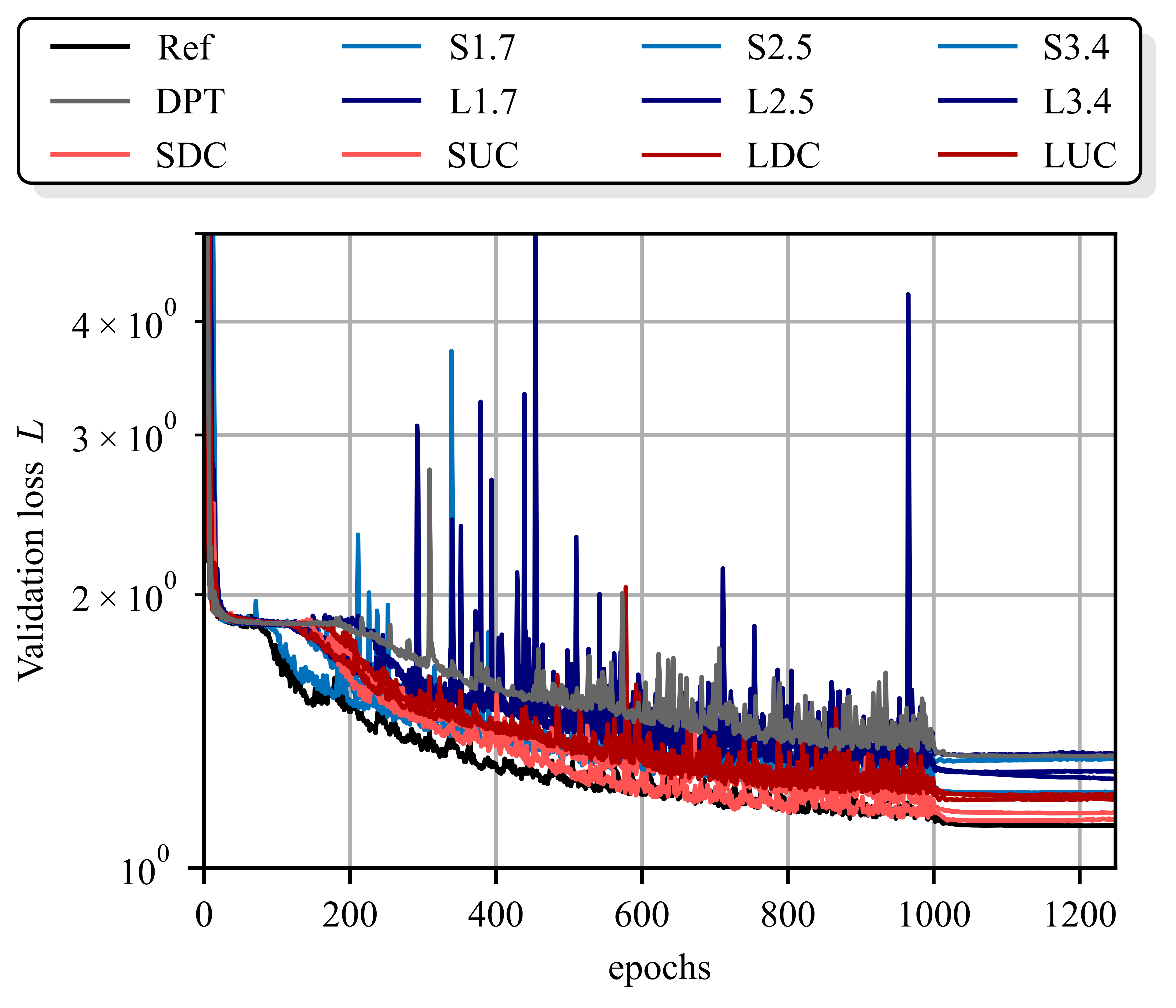}
    \caption{\textbf{Validation loss as a function of training epochs.}}
    \label{S1:TrainingCurves}
\end{figure}

Networks were trained for 1250 epochs with an Adam learning rate optimizer. Fig.~\ref{S1:TrainingCurves} shows the validation loss during training. The lines can be linked to the models by color.

\subsection{Precision and recall}

\begin{figure}
    \centering
    \includegraphics{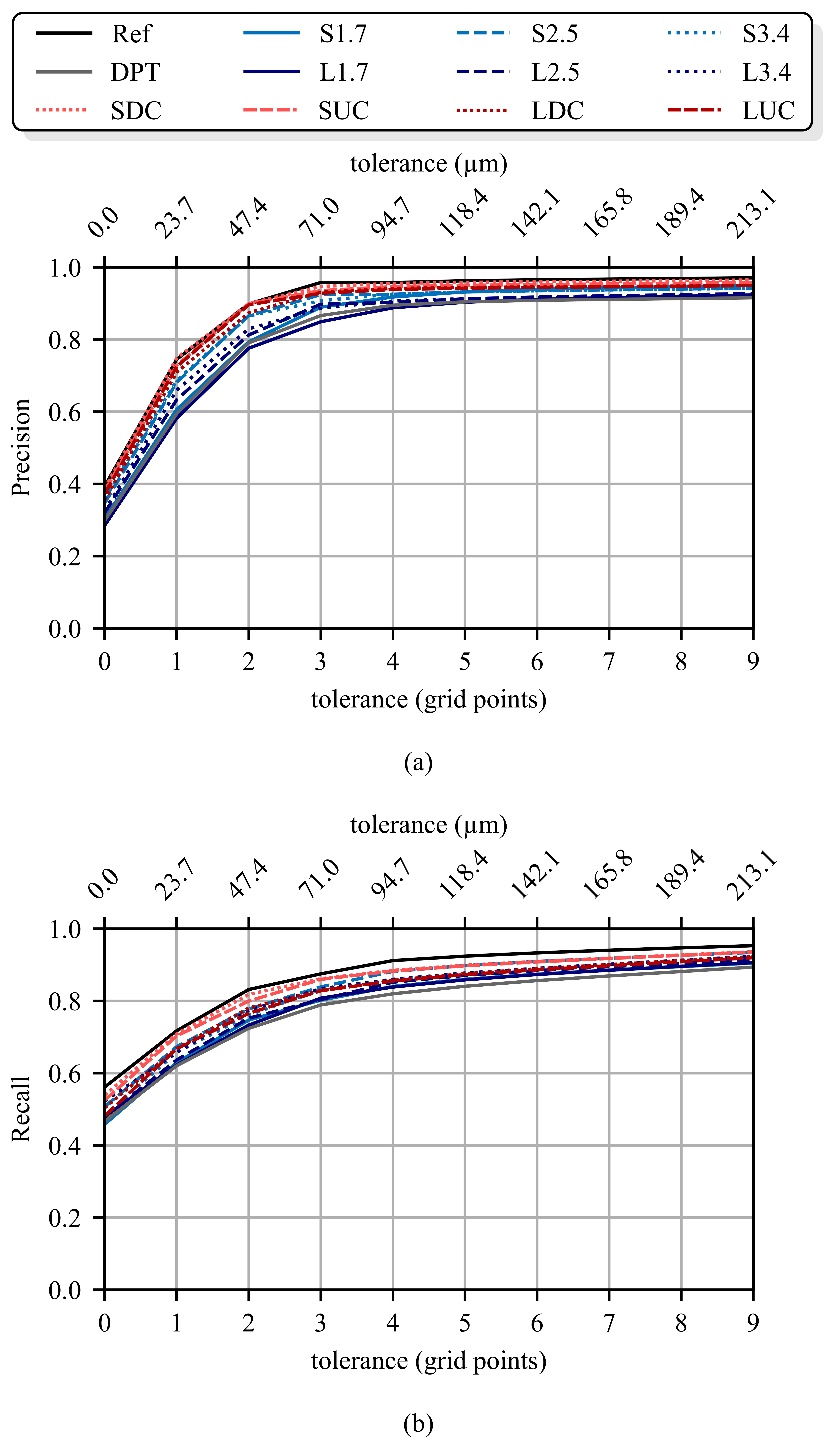}
    \caption{\textbf{Precision and recall over tolerance.} (a) Precision, (b) recall.}
    \label{fig:Precision_and_Recall}
\end{figure}

As described in the Methods section, the $F_1$ scores analyzed in the Results section are the harmonic mean of the precision and the recall. To provide a comprehensive evaluation of the results, Fig.~\ref{fig:Precision_and_Recall} displays the precision and recall curves of the $F_1$ scores presented in Fig. 2. 

\subsection{Performance versus pressure and density}

\begin{figure*}[ht]
    \centering
    \includegraphics{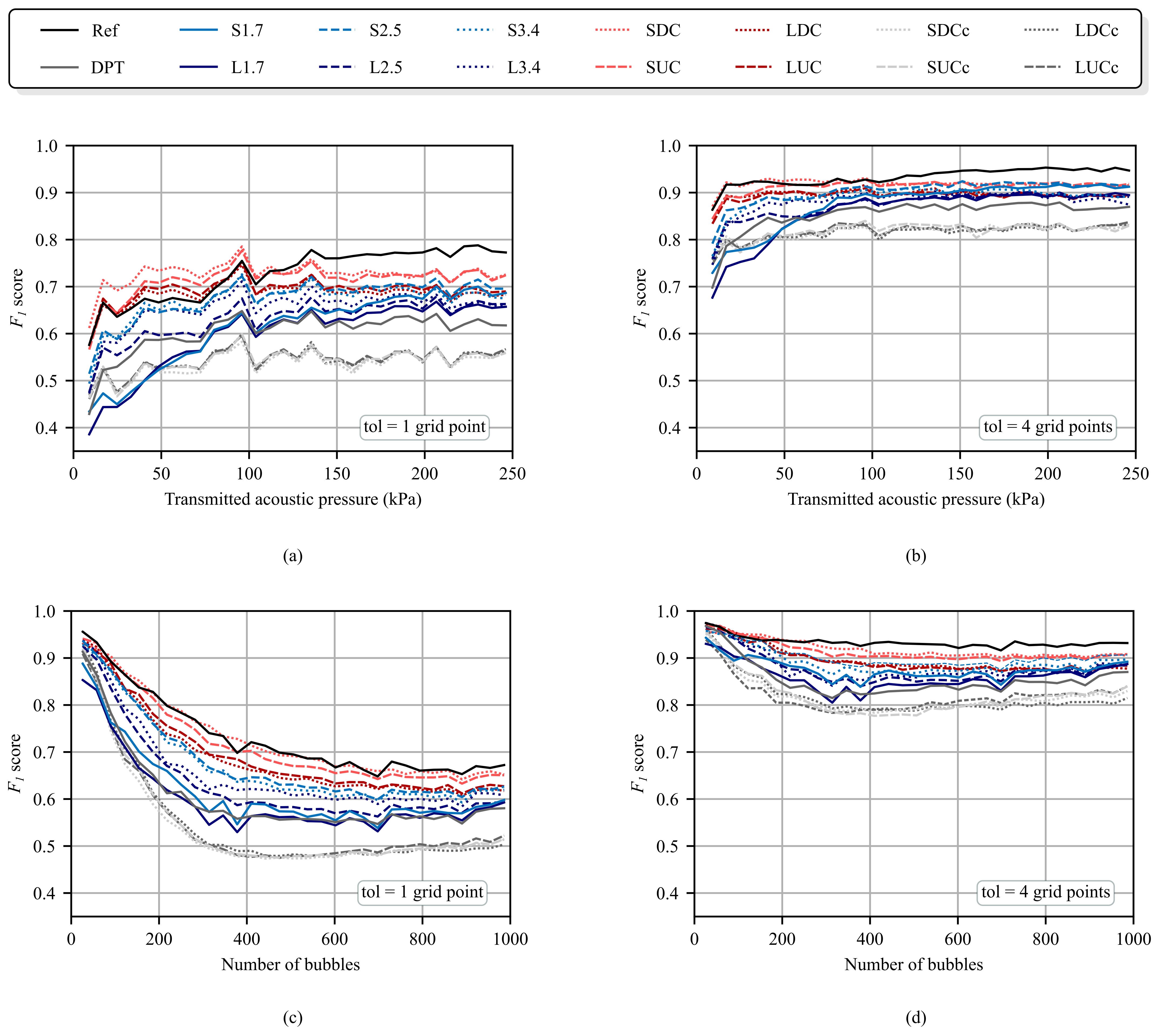}
    \caption{\textbf{Performance of the models within the training range for all models.} (a) and (b) show performance over acoustic pressure, (c) and (d) show performance over the number of bubbles.}
    \label{S1:PA_NB}
\end{figure*}

The relation between performance, acoustic pressure, and bubble density was analyzed for all models trained on raw pulses. For clarity, only a selection is displayed in the core of the paper. Fig. \ref{S1:PA_NB} displays the results for all the waveforms investigated.

\subsection{Image intensity}

\begin{wrapfigure}{l}{0.5\textwidth}
    \centering
    \includegraphics[width=0.9\linewidth]{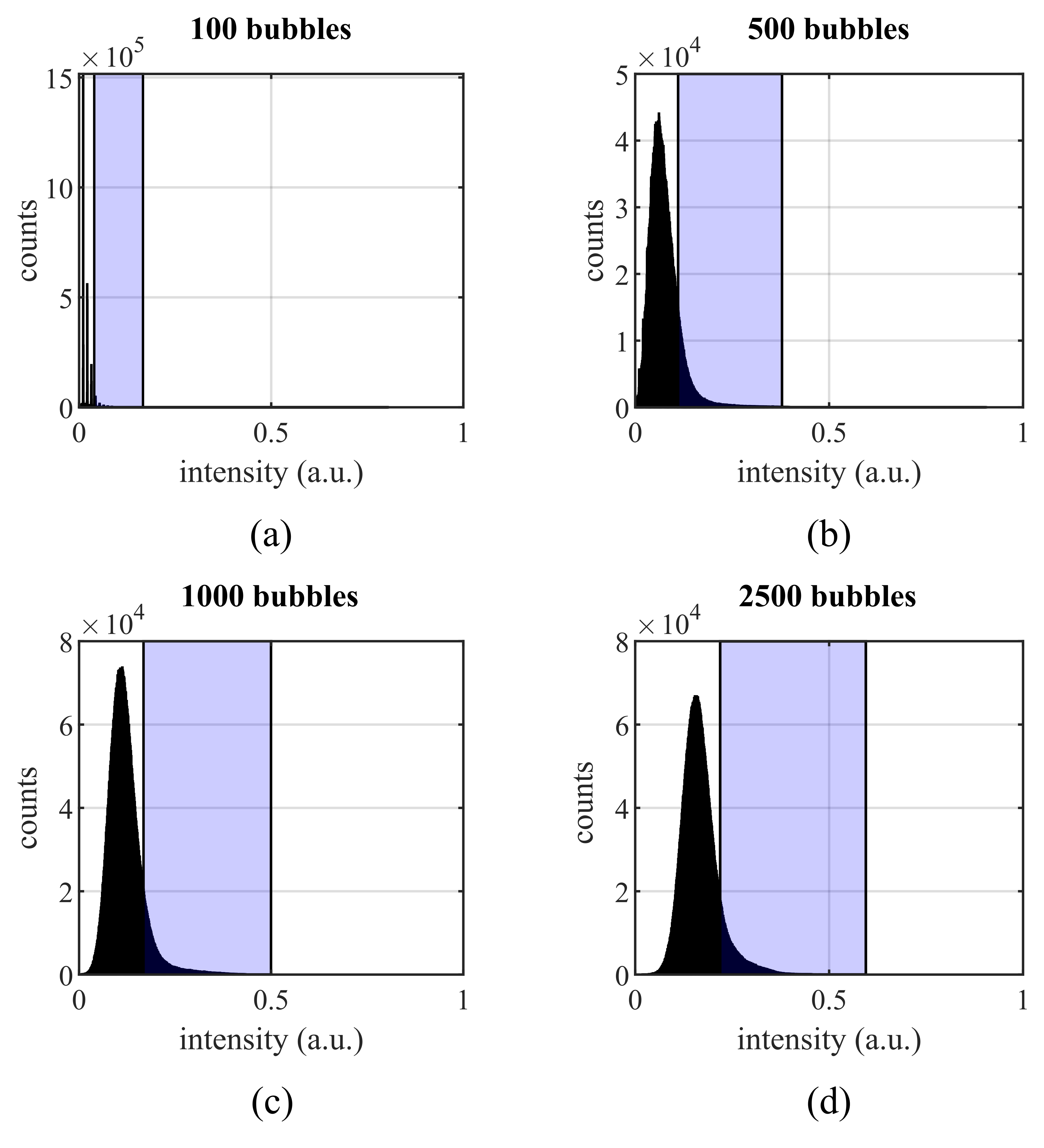}
    \caption{\textbf{Pixel intensity distributions of Fig.~\nBubblesComparison(e) to (h). }Pixel intensities lower than 0.001 are neglected. The blue patches display the range of the used colorbar.}
    \label{fig:intensity_histogram}
\end{wrapfigure}

Higher bubble densities automatically result in denser network predictions, and thus in a higher background clutter in the images. The histograms in Fig. \ref{fig:intensity_histogram} show the pixel intensity distribution of the images of which a section is displayed in Fig. 5(e) to (h), i.e., for an increasing number of bubbles. This shift in pixel intensity distribution complicates the direct comparison between different models and different bubble densities with the same color map. Therefore, for a fair comparison, we adapt the limits of the color bar using the mean ($\mu$) and standard deviation ($\sigma$) of the pixel intensity distributions. Pixel intensities below 0.001 are neglected in computing $\mu$ and $\sigma$ because they correspond to pixels where no bubble is predicted and they fall aside the main distribution. The lower and upper limit of the color bars are set to $\mu + \sigma$ and $\mu + 8\sigma$, respectively, to automatically set the dynamic range of the images.

\subsection{Considerations on robustness}

In Section III-C, we have assessed the performance of a series of models trained and evaluated on the same noise level. These results describe how robust the super-resolution pipeline, including the pulse selection, is to increasing noise levels. To further test robustness, one can investigate how individual networks perform across the noise levels specified in Table~\ref{tab:noise_list}. To explore this approach, we train two categories of networks.

\begin{figure*}[ht]
    \centering
    \includegraphics{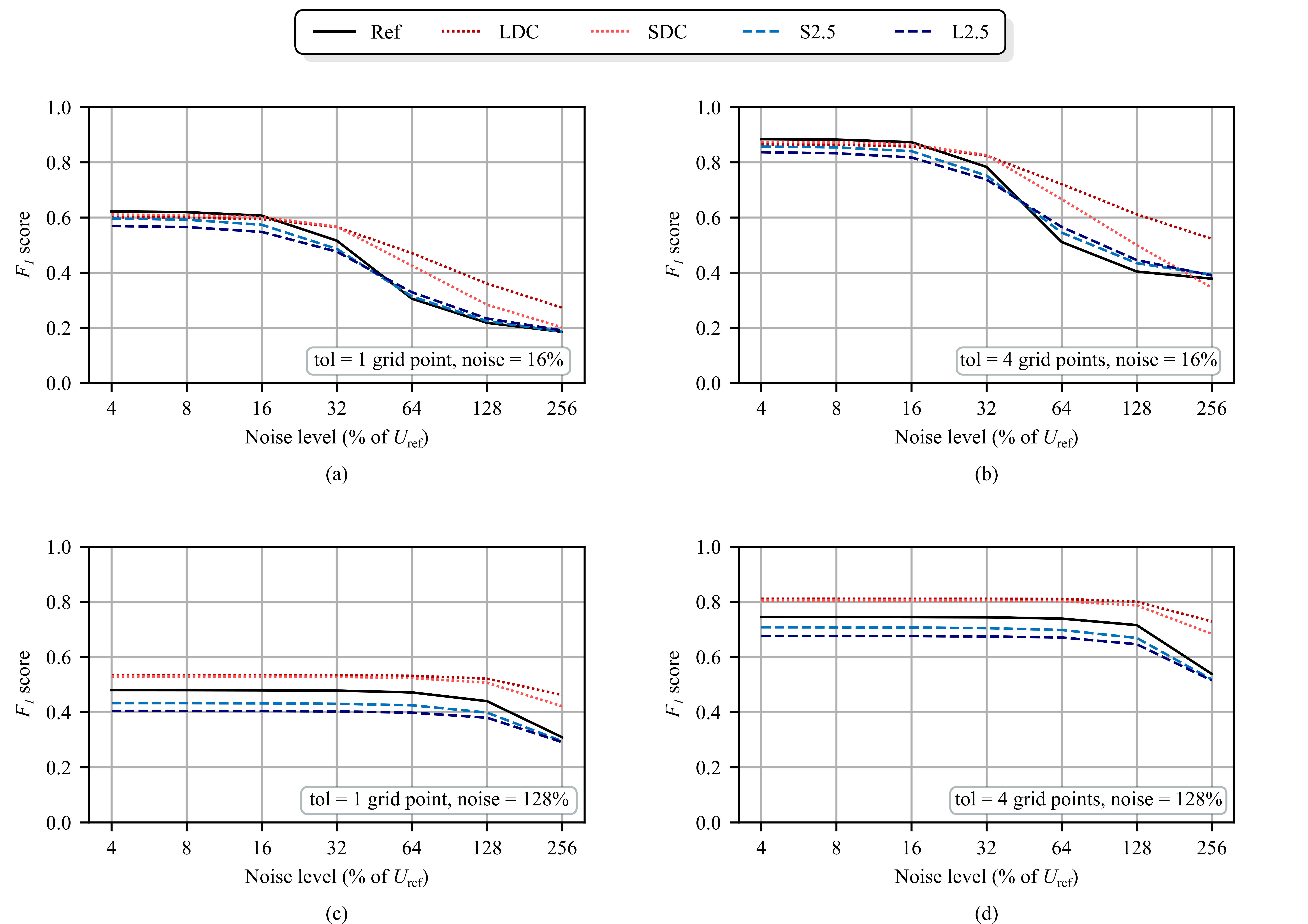}
    \caption{\textbf{Performance of networks trained on noise levels of 16\% and 128\% of $U_\mathrm{ref}$.}}
    \label{fig:robustness_singlenoise}
\end{figure*}

First, we train the evaluated networks on a single noise level and evaluate their performance across noise levels. Fig.~\ref{fig:robustness_singlenoise} displays the performance of networks trained on 16\% and 128\% of the noise level $U_\mathrm{ref}$, and for a tolerance of 1 and 4 grid points. The localization performance remains stable up to the noise level on which it is trained and drops substantially thereafter.

\begin{figure*}[ht]
    \centering
    \includegraphics{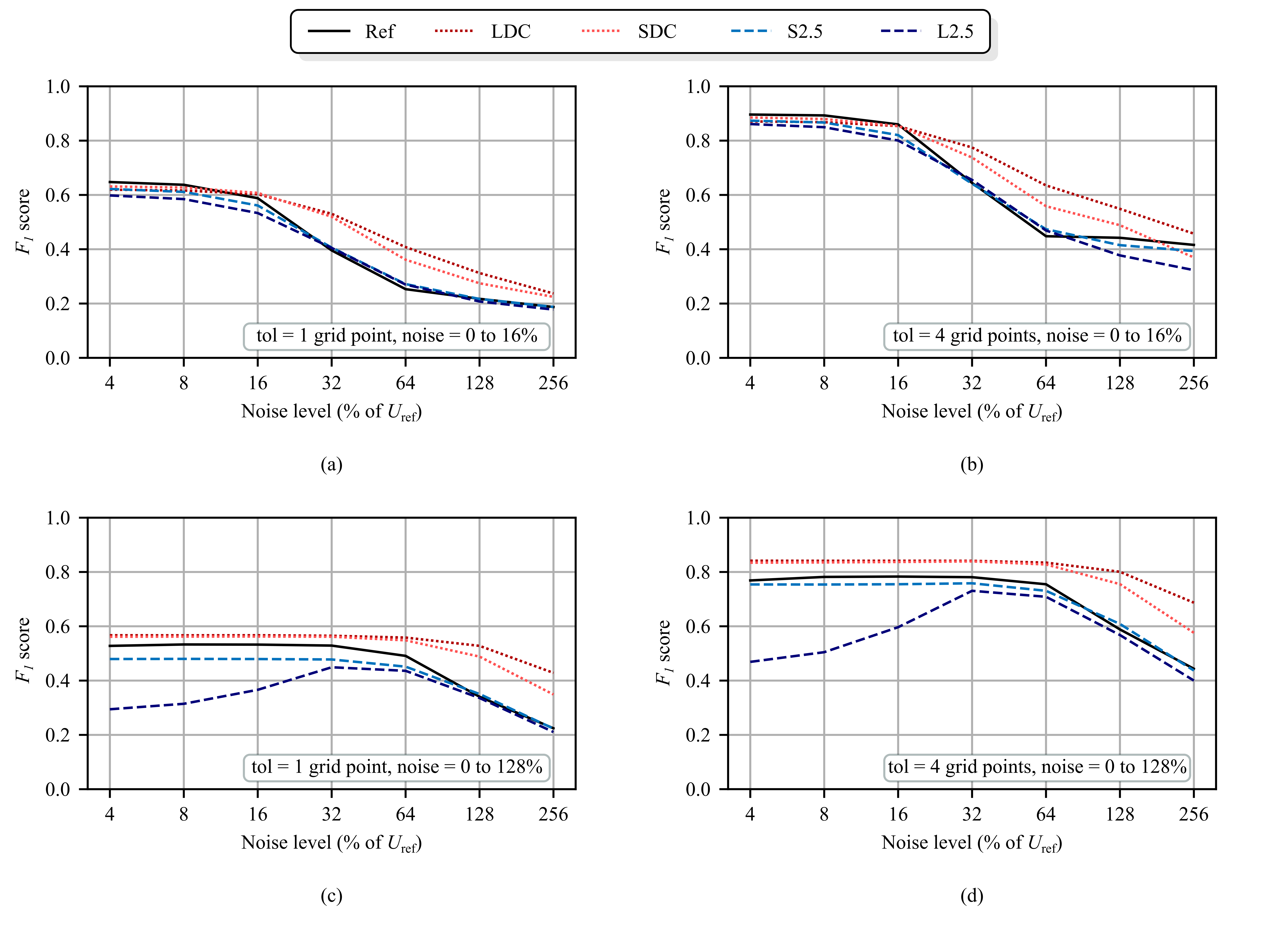}
    \caption{\textbf{Performance of networks trained on noise levels of 0\% to 16\% and 0\% to 128\% of $U_\mathrm{ref}$.}}
    \label{fig:robustness_trainingrange}
\end{figure*}

Second, we train the evaluated networks on a range of noise. The results displayed in Fig.~\ref{fig:robustness_trainingrange} present the performance of networks trained on 0\% to 16\% and 0\% to 128\% of $U_\mathrm{ref}$, and for tolerances of 1 and 4 grid points. 

In both Fig.~\ref{fig:robustness_singlenoise} and Fig.~\ref{fig:robustness_trainingrange} all other parameters are kept constant for fair comparison. The differences in performance between both approaches are small, except for the network trained on L2.5 with a noise range from 0\% to 128\%. Why this model shows a lower performance at the lower noise levels remains to be investigated. The approach evaluated in Section III-C is the most stable and best-performing approach regarding $F_1$ scores. Therefore, we believe that future efforts in optimizing deconvolutional super-resolution strategies should focus on pulse selection and tailoring the training parameters to specific conditions, such as SNR, rather than enhancing the performance generalization of individual networks.

\newpage

\begin{table}[h]
    \centering
    \caption{Noise values}
    \begin{tabular}{cc}
    \\\toprule
    \textbf{Noise level (\%)} & \textbf{Noise level (dB)} \\
    \midrule
    4 & -28.0\\
    8 & -21.9\\
    16 & -15.9\\
    32 & -9.9\\
    64 & -3.9\\
    128 & 2.1\\
    \bottomrule
    \end{tabular}\\
    Noise level is expressed as a fraction of $U_\mathrm{ref}$ in both percentage and dB.
    \label{tab:noise_list}
\end{table}

\end{document}